\documentclass[oldversion]{aa}
\usepackage{amsmath,graphicx,natbib,txfonts,longtable}

\begin{document}

\title{Evolution of the Early-Type Galaxy Fraction in Clusters since $z$ = 0.8\thanks{Based on observations obtained in visitor and service modes at the ESO Very Large Telescope (VLT) as part of the Large Programme 166.A-0162 (the ESO Distant Cluster Survey). Also based on observations made with the NASA/ESA  Hubble Space Telescope, obtained at the Space Telescope Science Institute, which is operated by the Association of Universities for Research in Astronomy, Inc., under NASA contract NAS 5-26555. These observations are associated with proposal 9476. Support for this proposal was provided by NASA through a grant from the Space Telescope Science Institute. }}

\author{Luc Simard \inst{1}, Douglas Clowe \inst{2}, Vandana Desai \inst{3}, Julianne J. Dalcanton \inst{4}, Anja von der Linden\inst{5,6}, Bianca M. Poggianti\inst{7}, Simon D. M. White \inst{6}, Alfonso Arag\'on-Salamanca\inst{8}, Gabriella De Lucia\inst{6, 9}, Claire Halliday\inst{10}, Pascale Jablonka\inst{11}, Bo Milvang-Jensen\inst{12,13}, Roberto P. Saglia\inst{14}, Roser Pell\'o \inst{15}, Gregory H. Rudnick \inst{16,17}, Dennis Zaritsky\inst{18}}

\institute{National Research Council of Canada, Herzberg Institute of Astrophysics, 5071 West Saanich Road, Victoria, British Columbia, Canada
\and
Ohio University, Department of Physics and Astronomy, Clippinger Lab 251B, Athens, OH 45701, USA
\and
California Institute of Technology, MS 320-47, Pasadena, CA 91125, USA
\and
University of Washington, Department of Astronomy, Box 351580, Seattle, WA 98195-1580, USA
\and
Kavli Institute for Particle Astrophysics and Cosmology, PO Box 20450, MS 29, Stanford, CA 94309, USA
\and
Max-Planck-Institut f\"ur Astrophysik, Karl-Schwarschild-Str., 1, Postfach 1317, D-85741 Garching
\and
INAF - Astronomical Observatory of Padova, Italy
\and
School of Physics and Astronomy, University of Nottingham, Nottingham, NG7 2RD, UK 
\and
INAF - Astronomical Observatory of Trieste, via Tiepolo 11, I-34143 Trieste, Italy
\and
INAF - Osservatorio Astronomico di Arcetri, Largo Enrico Fermi 5, 50125 Firenze, Italy
\and
Observatoire de l'Universit\'e de Gen\`eve, Laboratoire d'Astrophysique de l'Ecole Polytechnique F\'ed\'erale de Lausanne (EPFL), 1290 Sauverny, Switzerland
\and
Dark Cosmology Centre, Niels Bohr Institute, University of Copenhagen, Juliane Maries Vej 30, DK-2100 Copenhagen, Denmark
\and
The Royal Library / Copenhagen University Library, Research Department, Box 2149, DK-1016 Copenhagen K, Denmark
\and
Max-Planck Institut f\"ur extraterrestrische Physik, Giessenbachstrasse D-85748 Garching, Germany
\and 
Laboratoire d'Astrophysique de Toulouse-Tarbes, CNRS, Universit\'e de Toulouse, 14 Avenue Edouard Belin, 31400-Toulouse, France
\and
The University of Kansas, Department of Physics and Astronomy, Malott room 1082, 1251 Wescoe Hall Drive, Lawrence, KS, 66045, USA
\and
NOAO, 950 North Cherry Avenue, Tucson, AZ 85719, USA
\and 
Steward Observatory, University of Arizona, 933 North Cherry Avenue, Tucson, AZ, 85721
}

\date{Accepted for publication in A\&A }

\abstract{We study the morphological content of a large
sample of high-redshift clusters to determine its dependence on cluster
mass and redshift. Quantitative morphologies are based on PSF-convolved, 2D bulge+disk
decompositions of cluster and field galaxies on deep Very Large
Telescope FORS2 images of eighteen, optically-selected galaxy clusters at $0.45 < z < 0.80$ observed as part
of the ESO Distant Cluster Survey (``EDisCS''). Morphological content is characterized by the early-type
galaxy fraction $f_{et}$, and early-type galaxies are objectively selected based on their bulge fraction and image
smoothness. This quantitative selection is equivalent to selecting galaxies visually classified as E {\it or} S0. Changes in early-type fractions as a function of
cluster velocity dispersion, redshift and star-formation activity are
studied. A set of 158 clusters extracted from the Sloan
Digital Sky Survey is analyzed exactly as the distant EDisCS sample to
provide a robust local comparison. We also compare our results to a set of clusters from the Millennium Simulation. Our main results are: (1) The early-type fractions of the SDSS and EDisCS clusters exhibit no clear trend as a function of  cluster velocity dispersion. (2) Mid-$z$ EDisCS clusters around $\sigma$ = 500 km/s have $f_{et} \simeq$ 0.5 whereas high-$z$ EDisCS clusters have $f_{et} \simeq$ 0.4. This represents a $\sim$25$\%$ increase over a time interval of 2 Gyrs. (3) There is a marked difference in the morphological content of EDisCS and SDSS clusters. None of the EDisCS clusters have early-type galaxy fractions greater than 0.6 whereas half of the SDSS clusters lie above this value.  This difference is seen in clusters of all velocity dispersions. (4) There is a strong and clear correlation between morphology and star formation activity in SDSS and EDisCS clusters in the sense that decreasing fractions of [OII] emitters are tracked by increasing early-type fractions.  This correlation holds independent of cluster velocity dispersion and redshift even though the fraction of [OII] emitters decreases from $z \sim0.8$ to $z \sim 0.06$ in all environments. Our results pose an interesting challenge to structural transformation and star formation quenching processes that strongly depend on the global cluster environment (e.g., a dense ICM) and suggest that cluster membership may be of lesser importance than other variables in determining galaxy properties. 

}

\authorrunning{Simard et al.}
\titlerunning{Evolution of the Early-Type Galaxy Fraction in Clusters}

\maketitle

\section{Introduction}\label{intro}

Our current paradigm for the origin of galaxy morphologies rests upon
hierarchical mass assembly \citep[e.g.,][]{steinmetz02}, and many
transformational processes are at work throughout the evolutionary
histories of galaxies. Some determine the main structural traits
(e.g., disk versus spheroid) while others only influence properties
such as color and star-formation rates. Disk galaxy collisions lead to
the formation of elliptical galaxies
\citep{spitzer51,toomre72,farouki82,negroponte83,barnes92,barnes96,mihos96}, and the extreme example of
this process is the build-up of the most massive galaxies in the
Universe at the cores of galaxy clusters through the accretion of
cluster members. Disks can also be transformed into spheroidals by
tidal shocks as they are harassed by the cluster gravitational
potential \citep{farouki81,moore96,moore98}. Harassment inflicts more damage to low luminosity galaxies because of their slowly rising rotation curves and their low density cores. Galaxies can be stripped of their
internal gas and external supply through ram pressure exerted by the
intracluster medium \citep{gunn72,larson80,quilis00}, and the result is a
``quenching'' (or ``strangulation'') of their star formation that
leads to a rapid reddening of their colours \citep[also see][]{martig09}. The task of isolating observationally the effects of a given process has remained a major challenge to this day.

Many processes affecting galaxy morphologies are clearly
environmentally-driven, and galaxy clusters are therefore ideal
laboratories in which to study all of them. The dynamical state of a
cluster, which can be observationally characterized by measuring mass
and substructures, should be related to its morphological content. For
example, the number of interactions/collisions suffered by a given
galaxy should depend on local number density and the time it has spent
within the cluster. Dynamically young clusters with a high degree of
subclustering should contain large numbers of galaxies that are
infalling for the first time. More massive clusters will contain more
galaxies, but they will also have higher galaxy-galaxy relative
velocities that may impede merging
\citep{lubin02}. Spheroidal/elliptical galaxies will preferentially be
formed in environments where the balance between number density and
velocity dispersions is optimal, but it is still not clear where this
optimal balance lies. Cluster masses can be estimated
from their galaxy internal velocity dispersion
\citep{rood72,dressler84,carlberg97,tran99,borgani99,lubin02}, through weak-lensing
shear \citep{kaiser93,schneider95,hoekstra00,clowe06} or through analysis of their hot X-ray emitting atmospheres \citep[e.g., ][]{allen98}, and it will be
used here as the main independent variable against which morphological
content will be studied.

The morphological content of high-redshift clusters is most often
characterized by the fraction $f_{E+S0}$ of early-type galaxies they
contain \citep{dressler97,dokkum00, fasano00, dokkum01, lubin02, holden04,
smith05, postman05, desai07,poggianti09b}. The bulk of the data available so far is based on
visual classification. ``Early-type'' galaxies are defined in terms of
visual classifications as galaxies with E or S0 Hubble types. A
compilation of early-type fractions taken from the literature
\citep{dokkum00} shows a dramatic increase of the early-type fractions
as a function of decreasing redshift from values around 0.4$-$0.5 at
$z \sim 1$ to values around 0.8 in the local Universe. However, the
interpretation of this trend is not entirely clear as others
\citep[e.g.,][]{dressler97,fasano00,desai07,poggianti09b} have reported that the fraction of E's
remains unchanged as a function of redshift and that the observed
changes in early-type fractions are entirely due to the S0 cluster
populations. S0 populations were observed to grow at the expense of
the spiral population \citep{smith05,postman05,moran07,poggianti09b} although others \citep[e.g.,][]{holden09} have argued for no evolution in the relative fraction of ellipticals and S0s with redshift. \citet{smith05} and \citet{postman05} show that
the evolution of $f_{E+S0}$ is in fact a function of both lookback
time (redshift) and projected galaxy density. They find $f_{E+S0}$
stays constant at 0.4 over the range 1 $< t_{lookback} <$ 8 Gyr for
projected galaxy densities $\Sigma <$ 10 Mpc$^{-2}$. For high density
environments ($\Sigma$ = 1000 Mpc$^{-2}$), $f_{E+S0}$ decreases from
0.9 to 0.7. At fixed lookback time, $f_{E+S0}$ varies by a factor of
1.8 from low to high densities at $t_{lookback}$ = 8 Gyr and by a
factor of 2.3 at $t_{lookback}$ = 1 Gyr. The difference between low
and high density environments thus increases with decreasing lookback
time. Both studies indicate that the transition between low and high
densities occurs at $0.6R_{200}$ ($R_{200}$ is the projected radius delimiting a sphere with interior mean density 200 times the critical density at the cluster redshift, see Equation~\ref{radius200}). \citet{postman05} also find that
$f_{E+S0}$ does not change with cluster velocity dispersion for
massive clusters ($\sigma$ $>$ 800 km/s). The data for one of their clusters also suggest that
$f_{E+S0}$ decreases for lower mass systems.  This trend would be
consistent with observations of $f_{E+S0}$ in groups that show a
strong trend of decreasing $f_{E+S0}$ versus decreasing $\sigma$
\citep{zabludoff98}. Finally, $f_{E+S0}$ seems to correlate with cluster X-ray luminosity at the 2-3$\sigma$ level \citep{postman05}.

Recent works on stellar mass-selected cluster galaxy samples \citep{holden07,vanderwel07} paint a different picture. The fractions of E+S0 galaxies in clusters, groups and the field do not appear to have changed significantly from $z \sim 0.8$ to $z \sim 0.03$ for galaxies with masses greater than 4$\times 10^{10} M_{\odot}$. The mass-selected early-type fraction remains around 90\% in dense environments ($\Sigma >$ 500 gal Mpc$^{-2}$) and 45\% in groups and the field. These results show that the morphology-density relation of galaxies more massive than 0.5M$_{*}$ has changed little since $z \sim 0.8$ and that the trend in morphological evolution seen in luminosity-selected samples must be due to lower mass galaxies. This is in agreement with \citet{delucia04,delucia07} and \citet{rudnick09} who have shown the importance of lower mass (i.e., fainter) galaxies to the evolution of the color-magnitude relation and of the luminosity function versus redshift. Another interesting result has come from attempts to disentangle age, morphology and environment in the Abell 901/902 supercluster \citep{wolf07,lane07}. Local environment appears to be more important to galaxy morphology than global cluster properties, and while the expected morphology-density and age-morphology relations have been observed, there is no evidence for a morphology-density relation at a fixed age. The time since infall within the cluster environment and not density might thus be the more fundamental parameter dictating the morphology of cluster galaxies.

A number of efforts have been made on the theoretical side to model the
morphological content of clusters. \citet{diaferio01} used a model in
which the morphologies of cluster galaxies are solely determined by
their merger histories. A merger between two similar mass galaxies
produces a bulge, and a new disk may form through the subsequent
cooling of gas. Bulge-dominated galaxies are in fact formed by mergers
in smaller groups that are later accreted by clusters. Based on their
model, they reach the following conclusions: (1) the fraction of
bulge-dominated galaxies inside the virial radius should depend on the
mass of the cluster, and it should show a pronounced
peak for clusters with mass of 3 $\times$ 10$^{14}$ M$_\odot$ followed
by a decline for larger cluster masses. (2) The fraction of
bulge-dominated galaxies should be independent of redshift for
clusters of fixed mass, and (3) the dependence of morphology on
cluster mass should be stronger at high redshift than at low
redshift. \citet{lanzoni05} use the GALICS semi-analytical models and
find that early-type fractions strongly depend on galaxy luminosity
rather than cluster mass.  By selecting a brighter subsample of
galaxies from their simulations, they find a higher fraction of
ellipticals irrespective of the cluster mass in which these galaxies
reside. This trend is particularly noticeable in their high-density
environments. Observations and these earlier models clearly do not agree in
important areas, and a comparison between them would clearly benefit
from a larger cluster sample size. More recently, the Millennium Simulation \citep[MS;][]{springel05} has provided the highest resolution model thus far of a large (0.125 Gpc$^3$), representative volume  of the Universe. Improved tracking of dark matter structure and new semi-analytical prescriptions \citep{delucia07a} allow the evolution of the galaxy population to be followed with higher fidelity and better statistics than in the otherwise similar work of \citet{diaferio01}. We will use cluster catalogues from the MS later in this paper for comparison with our observational data. 

Our understanding of high-redshift cluster galaxy populations in terms
of their evolution as a function of redshift and their
cluster-to-cluster variations has been hampered by the lack of
comprehensive multi-wavelength (optical, near-infrared and X-ray)
imaging and spectroscopic studies of large, homogeneously-selected
samples of clusters. Many efforts are underway to improve sample sizes
\citep{gonzalez01,gladders05,willis05,postman05}. One of these efforts
is the European Southern Observatory Distant Cluster Survey
\citep[``EDisCS'';][]{white05}. The EDisCS survey is an ESO large
programme aimed at the study of a sample of eighteen
optically-selected clusters over the redshift range 0.5-0.8. It makes
use of the FORS2 spectrograph on the Very
Large Telescope for optical imaging and spectroscopy and of the SOFI imaging
spectrograph on the New Technology Telescope (NTT) for near-infrared
imaging. A number of papers on star formation in clusters \citep{poggianti06,poggianti09a} and the assembly of the cluster red sequence \citep{delucia04,delucia07,sanchez09,rudnick09} have been so far published from these data. In addition to the core VLT/NTT observations, a wealth of
ancillary data are also being collected. A 80-orbit program for the
Advanced Camera for Surveys (ACS) on the Hubble Space Telescope was
devoted to the $i$-band imaging of our ten highest-redshift
clusters. Details of the HST/ACS observations and visual galaxy
classifications are given in \citet{desai07} and the frequency and properties of galaxy bars is studied in \citet{barazza09}. X-ray observations with
the XMM-Newton satellite of three EDisCS clusters have been published
in \citet{johnson06} with more clusters being observed. H-alpha observations of three clusters have been
published in \citet{finn05} with more clusters also being
observed. Finally, the analysis of Spitzer/IRAC observations of all EDisCS clusters
is in progress (Finn et al., in preparation).

This paper presents the early-type galaxy fractions of EDisCS clusters
as a function of cluster velocity dispersion, redshift and
star-formation activity.  A set of local clusters extracted from
the Sloan Digital Sky Survey (SDSS) is used as a comparison
sample. Early-type fractions were measured from
two-dimensional bulge+disk decompositions on deep, optical VLT/FORS2
and HST/ACS images of spectroscopically-confirmed cluster member
galaxies. Section~\ref{data} describes the EDisCS cluster sample
selection and the imaging data. Section~\ref{analysis} describes the
procedure used to perform bulge+disk decompositions on SDSS, VLT/FORS2
and HST/ACS images. Section~\ref{efractions} presents early-type
fractions for the EDisCS clusters with a detailed comparison between
visual and quantitative morphologies and between HST- and VLT-derived
early-type fractions. It also includes early-type fractions for the
SDSS clusters. Changes in EDisCS early-type fractions as a function of 
cluster velocity dispersion, redshift and star-formation activity are
studied in Section~\ref{results}. Finally, Sections~\ref{discussion}
and~\ref{conclusions} discuss our results and their implications for
the morphological content of clusters. The set of cosmological
parameters used throughout this paper is ($H_0, \Omega_{m},
\Omega_{\Lambda}$) = (70, 0.3, 0.7).

\section{Data}\label{data}

\subsection{Sample Selection and VLT/FORS2 Optical Imaging}\label{sampsel}
The sample selection and optical/near-infrared imaging data for the
EDisCS survey are described in details in \citet{gonzalez02},
\citet{white05} (optical photometry) and Arag\'on-Salamanca et
al. (near-IR photometry; in preparation). Photometric redshifts
for the EDisCS clusters are presented in \citet{pello09}, and cluster velocity dispersions measured from
weak-lensing mass reconstructions are given in
\cite{clowe06}. Spectroscopy for the EDisCS clusters is detailed in
\citet{halliday04} and \citet{milvang08}. Clusters in the EDisCS
sample were drawn from the Las Campanas Distant Cluster Survey (LCDCS)
candidate catalog \citep{gonzalez01}. Candidate selection was
constrained by published LCDCS redshift and surface brightness
estimates.  Candidates were selected to be among the highest surface
brightness detections at each redshift in an attempt to recover some
of the most massive clusters at each epoch. Using the estimated
contamination rate for the LCDCS of $\sim 30 \%$, we targeted thirty
candidates in the redshift range 0.5$-$0.8 for snapshot VLT/FORS2
imaging in an effort to obtain twenty (10 at $z \sim 0.5$ and 10 at $z
\sim 0.8$) confirmed clusters.

The $z \sim 0.5$ candidates were observed for 20 minutes in each of
$I_\mathrm{B}$ and $V_\mathrm{B}$, and the $z\sim0.8$ candidates were
observed for 20 minutes in each of $I_\mathrm{B}$ and
$R_\mathrm{sp}$. These filters are the standard FORS2 ones.  $V_\mathrm{B}$ and $I_\mathrm{B}$ are close approximations to the \citet{bessell90} photometric system while the $R_\mathrm{sp}$ is a special filter for FORS2. Final cluster candidates for deeper VLT imaging were
selected on the basis of color and surface density of galaxies on the
sky \citep{white05}. The image quality on the final stacked images
ranged from 0\farcs 4 to 0\farcs 8. As described in \citet{white05},
deep spectroscopy was not obtained for two cluster candidates
(1122.9-1136 and 1238.5-1144), and we therefore did not include them
here. The main characteristics (positions, redshifts, velocity
dispersions and radii) of the EDisCS cluster sample used in this paper
are given in Table~\ref{clsample}. $R_{200}$ is the projected radius
delimiting a sphere with interior mean density 200 times the critical
density at the cluster redshift, and it is used throughout this paper
as an important fiducial radius. $R_{200}$ values in
Table~\ref{clsample} were calculated using the equation :

\begin{eqnarray}
R_{200} = 1.73 \frac{\sigma}{1000 {\rm km/s}} \frac{1}{\sqrt{\Omega_\Lambda + \Omega_m(1+z)^3}} h_{100}^{-1} {\rm Mpc}
\label{radius200}
\end{eqnarray}

\noindent where $h_{100}$ = $H_0$ / 100 and $\sigma_{cluster}$ is the cluster velocity dispersion measured using spectroscopically-confirmed cluster members \citep{carlberg97,finn05}. Cluster masses were calculated using the equation:

\begin{eqnarray}
M_{cl} = 1.2\times10^{15} \biggl (\frac{\sigma}{1000 {\rm km/s}}\biggr )^3\frac{1}{\sqrt{\Omega_\Lambda + \Omega_m(1+z)^3}} h_{100}^{-1} M_\odot
\label{clmass}
\end{eqnarray}

\noindent as in \citet{finn05}. 

In practice, the redshift distributions of high-$z$ and the mid-$z$ samples partly overlap as can be seen from Table~\ref{clsample}. 

\begin{table*}
\caption[]{Main characteristics of the EDisCS cluster sample: IDs, positions, redshifts, number of spectroscopically-confirmed members, velocity dispersions and radii. Clusters with HST imaging are identified by the superscript ``h'' in their ID.}
\begin{center}
\begin{tabular*}{15.5cm}{ccccccrcc}
Mid-$z$ clusters & & & & & & & &\\
\hline
ID & RA$^{\rm a}$ & DEC$^{\rm a}$ & z$^{\rm b}$ & Age of Universe  & $N_{mem}$$^{\rm c}$ & $\sigma$$^{\rm d}$ & $R_{200}$$^{\rm e}$  & $M_{cl}$$^{\rm f}$ \\
& (2000.0) & (2000.0) & & ($\times$ $t_0$) & & (km/s) & (Mpc) & (10$^{15}$M$_{\odot}$)\\
(1) & (2) & (3) & (4) & (5) & (6) & (7) & (8) & (9)\\
\hline
1018.8-1211 & 10:18:46.8 & $-$12:11:53 & 0.4716 & 0.654 & 33 & 474 $^{+~75}_{-~57}$ & 0.91 & 0.142\\
1059.1-1253 & 10:59:07.1 & $-$12:53:15 & 0.4550 & 0.663 & 41 & 517 $^{+~71}_{-~40}$ & 1.00 & 0.186\\
1119.3-1130 & 11:19:16.7 & $-$11:30:29 & 0.5491 & 0.615 & 21 & 165 $^{+~34}_{-~19}$ & 0.30 & 0.006\\
1202.7-1224 & 12:02:43.4 & $-$12:24:30 & 0.4246 & 0.680 & 21 & 540 $^{+139}_{-~83}$ & 1.07 & 0.216\\
1232.5-1250$^h$ & 12:32:30.5 & $-$12:50:36 & 0.5419 & 0.618 & 54 & 1080 $^{+119}_{-~89}$ & 1.99 & 1.610\\
1301.7-1139 & 13:01:40.1 & $-$11:39:23 & 0.4828 & 0.648 & 37 & 681 $^{+~86}_{-~86}$ & 1.30 & 0.418\\
1353.0-1137 & 13:53:01.7 & $-$11:37:28 & 0.5889 & 0.596 & 22 & 663 $^{+179}_{-~91}$ & 1.19 & 0.362\\
1411.1-1148 & 14:11:04.6 & $-$11:48:29 & 0.5200 & 0.629 & 26 & 709 $^{+180}_{-105}$ & 1.32 & 0.461\\
1420.3-1236 & 14:20:20.0 & $-$12:36:30 & 0.4969 & 0.641 & 27 & 225 $^{+~77}_{-~62}$ & 0.43 & 0.015\\
\hline
& & & & & & & &\\
High-$z$ clusters & & & & & & & &\\
\hline
ID & RA & DEC & z & Age of Universe & $N_{mem}$ & $\sigma$ & $R_{200}$ & $M_{cl}$\\
& (2000.0) & (2000.0) & & ($\times$ $t_0$) & & (km/s) & (Mpc) & (10$^{15}$M$_{\odot}$)\\
(1) & (2) & (3) & (4) & (5) & (6) & (7) & (8) & (9)\\
\hline
1037.9-1243$^h$ & 10:37:51.2 & $-$12:43:27 & 0.5800 & 0.600 & 19 & 315 $^{+~76}_{-~37}$ & 0.57 & 0.039\\
1040.7-1156$^h$ & 10:40:40.4 & $-$11:56:04 & 0.7020 & 0.548 & 30 & 418 $^{+~55}_{-~46}$ & 0.70 & 0.085\\
1054.4-1146$^h$ & 10:54:24.5 & $-$11:46:20 & 0.6965 & 0.550 & 49 & 589 $^{+~78}_{-~70}$ & 0.99 & 0.238\\
1054.7-1245$^h$ & 10:54:43.6 & $-$12:45:52 & 0.7503 & 0.529 & 36 & 504 $^{+113}_{-~65}$ & 0.82 & 0.144\\
1103.7-1245b$^h$ & 11:03:36.5 & $-$12:44:22 & 0.7029 & 0.548 & 11 & 242 $^{+126}_{-104}$ & 0.40 & 0.016\\
1138.2-1133$^h$ & 11:38:10.3 & $-$11:33:38 & 0.4801 & 0.649 & 48 & 737 $^{+~77}_{-~56}$ & 1.41 & 0.531\\
1216.8-1201$^h$ & 12:16:45.1 & $-$12:01:18 & 0.7955 & 0.513 & 67 & 1018 $^{+~73}_{-~77}$ & 1.61 & 1.159\\
1227.9-1138$^h$ & 12:27:58.9 & $-$11:35:13 & 0.6375 & 0.575 & 22 & 572 $^{+~96}_{-~54}$ & 0.99 & 0.226\\
1354.2-1231$^h$ & 13:54:09.7 & $-$12:31:01 & 0.7562 & 0.527 & 21 & 668 $^{+161}_{-~80}$ & 1.08 & 0.335\\
\hline
\label{clsample}
\end{tabular*}

$^{\rm a}$ Cluster BCG Coordinates (J2000)\\
$^{\rm b}$ Cluster redshift measured from EDisCS spectroscopy\\
$^{\rm c}$ Number of cluster members confirmed by EDisCS spectroscopy\\
$^{\rm d}$ Cluster velocity dispersion measured from EDisCS spectroscopy\\
$^{\rm e}$ From equation~\ref{radius200}\\
$^{\rm f}$ From equation~\ref{clmass}\\
\end{center}
\end{table*}

\subsection{VLT Spectroscopy and Cluster Membership}\label{vltspec-members}

We use only spectroscopically-confirmed cluster members to calculate
our cluster early-type fractions. Deep multislit spectroscopy of the
EDisCS was obtained with the FORS2 spectrograph on VLT. Spectra of $>$
100 galaxies per cluster field were obtained with typical exposure times of
two and four hours for the mid-z and high-z samples
respectively. Spectroscopic targets were selected from $I$-band
catalogues. This corresponds to rest-frame $\sim$ 5000 $\pm$ 400 $\AA$
at the redshifts of the EDisCS clusters. Conservative rejection
criteria based on photometric redshifts were used in the selection of
spectroscopic targets to reject a significant fraction of non-members
while retaining a spectroscopic sample of cluster galaxies equivalent
to a purely $I$-band selected one. We verified {\it a posteriori} that
these criteria excluded at most 1$\%$ of the cluster galaxies
\citep{halliday04,milvang08}. The spectroscopic selection,
observations and spectroscopic catalogs are presented in detail in
\citet{halliday04} and \citet{milvang08}. As described in
\citet{halliday04}, cluster redshifts and velocity dispersions were
iteratively calculated using a biweight scale estimator for
robustness. Cluster members were defined as galaxies with redshifts
within the range $z_{cluster} \pm 3\sigma_{cluster}$ where
$z_{cluster}$ is the median redshift of all cluster members.
 
\subsection{HST/ACS Imaging}\label{hstimaging}

In addition to our ground-based imaging, a 80-orbit program (GO 9476,
PI: Dalcanton) for the Advanced Camera for Surveys (ACS) on the Hubble
Space Telescope (HST) was devoted to the $i$-band imaging of our ten
highest-redshift cluster fields. Details of these observations are
given in \citet{desai07}. Briefly, the HST observations were designed
to coincide as closely as possible with the coverage of the
ground-based optical imaging and spectroscopy, within guide star
constraints. The VLT/FORS2 images cover a 6\farcm5 $\times$ 6\farcm 5
region around each cluster, with the cluster center displaced by
1\arcmin from the center of the region. For reference, the ACS WFC has
a field of view of roughly 3\farcm 5 $\times$ 3\farcm 5. Balancing
scientific motives for going deep over the entire spectroscopic field
against a limited number of available orbits, we tiled each 6\farcm 5
$\times$ 6\farcm 5 field in four 1-orbit pointings overlapping one additional deep 4-orbit
pointing on the cluster center. The resulting exposure time per pixel
was 2040 seconds except for the central 3\farcm 5 $\times$ 3\farcm 5,
which had an exposure time per pixel of 10200 seconds. The deep
central pointing probes to lower surface brightness, fainter
magnitudes, and larger galactic radii in the region of the cluster
containing the most galaxies. All exposures were taken under LOW SKY
conditions to maximize our surface brightness sensitivity. An image
mosaic was created for each cluster using the CALACS/Multidrizzle
pipeline, and the final sampling of the multidrizzled image mosaics was 0\farcs045. This is the "native" ACS image sampling, and it was chosen to avoid potential aliasing problems that might have been introduced by a finer multidrizzle sampling given our limited dither pattern in the cluster outskirts. Clusters with HST imaging are identified by a ``h'' in
Table~\ref{clsample}.

\section{Quantitative Galaxy Morphology}\label{analysis}

\subsection{Source Detection and Extraction}\label{sources}

The source catalogs and segmentation images for the EDisCS clusters
were created using the SExtractor (``Source Extractor'') galaxy
photometry package version 2.2.2 \citep{bertin96}. The SExtractor
source detection was run on the combined deep FORS2 images in
``two-image'' mode using the I-band image as the reference detection
image for all the other passbands.  The detection threshold was
1.5$\sigma_{bkg}$, and the required minimum object area above that
threshold was 4 pixels.  The convolution kernel was a 7$\times$7
Gaussian kernel with a FWHM of 3.0 pixels.  No star/galaxy separation
based on the SExtractor ``stellarity'' index was attempted.  Every
source was fit with a bulge+disk model, and unresolved sources such as
stars could easily be identified as output models with zero half-light
radius.

As SExtractor performs source detection and photometry, it is able to
deblend sources using flux multi-thresholding.  This deblending
technique works well in the presence of saddle points in the light
profiles between objects.  Each SExtractor pre-deblending ``object''
consists of all the pixels above the detection threshold that are
spatially connected to one another.  This group of pixels may or may
not include several real objects.  The multi-thresholding algorithm
assigns the pixels between two adjacent objects and below the
separation threshold based on a probability calculated from bivariate
Gaussian fits to the two objects.  No assumption is made regarding the
shape of the objects in this statistical deblending technique. We used
a value for the SExtractor deblending parameter DEBLEND$_-$MINCONT of
0.0005. This value is {\it subjective}, and it was found through
visual inspection of several EDisCS cluster images to provide good
object separation.  Even though the value of DEBLEND$_-$MINCONT was
determined subjectively, it provides an unequivocal definition of an
object in the EDisCS catalogs.  It was only determined once, and the
same value of DEBLEND$_-$MINCONT was consistently used for all EDisCS
cluster images as well as for all the reliability tests of
Section~\ref{reliability}.

\subsection{Two-Dimensional Bulge+Disk Decompositions} \label{bdcomps}

This work uses GIM2D (Galaxy IMage 2D) version 3.2, a 2D decomposition fitting
program \citep{simard02}, to measure the structural parameters of
galaxies on the EDisCS VLT/FORS2 and HST/ACS images.  GIM2D is an
IRAF\footnote{IRAF is distributed by the National Optical Astronomy
Observatories, which are operated by the Association of Universities
for Research in Astronomy, Inc., under cooperative agreement with the
National Science Foundation.}/SPP package written to perform detailed
bulge+disk surface brightness profile decompositions of low
signal-to-noise (S/N) images of distant galaxies in a fully automated
way. GIM2D is publicly available, and it has been used extensively in
a wide range of different projects so far.

\subsubsection{Fitting Model} \label{fitmodel}

The fitting model used for the two-dimensional bulge+disk
decompositions of EDisCS galaxies is the same as the one used by
\citet{simard02}. It consists of a ``bulge'' component with a de
Vaucouleurs profile and of an exponential ``disk'' component. We put
``bulge'' and ``disk'' between quotes to emphasize that this
conventional nomenclature does does not say anything about the
internal kinematics of the components.  The presence of a ``disk''
component does not necessarily imply the presence of an actual disk
because many dynamically hot systems also have simple exponential
profiles. The fitting model had ten free parameters: the total galaxy
flux $F$, the bulge fraction $B/T$ ($\equiv$ 0 for pure disk systems),
the bulge semi-major axis effective radius $r_e$, the bulge
ellipticity $e$ ($e \equiv 1-b/a$, $b \equiv$ semi-minor axis, $a
\equiv$ semi-major axis), the bulge position angle of the major axis
$\phi_{b}$ on the image (clockwise, y-axis $\equiv$ 0), the disk
semi-major axis exponential scale length $r_d$ (also denoted $h$ in
the literature), the disk inclination $i$ (face-on $\equiv$ 0), the
disk position angle $\phi_d$ on the image, the subpixel $dx$ and $dy$
offsets of the model center with respect to the input science image
center. The sky background is not a free parameter of the fits (see
Section~\ref{skyestm}). The S\'ersic index for the bulge profile is
fixed at a value of $n = 4$ (i.e., the de Vaucouleurs profile value). The position angles $\phi_b$ and $\phi_d$
were not forced to be equal for two reasons: (1) a large difference
between these position angles is a signature of strongly barred galaxies, and
(2) some observed galaxies do have {\it bona fide} bulges that are not
quite aligned with the disk position angle. 

The smooth bulge+disk model used here is obviously a simple
approximation.  After all, many real galaxies will exhibit more than
two structural components such as nuclear sources, bars, spiral arms
and HII regions.  Even in the presence of only a bulge and a disk, the
ellipticity and/or the position angles of these components might be
functions of galactocentric distance. The bulge+disk model is a
trade-off between a reasonable number of fitting parameters and a
meaningful decomposition of distant galaxy images. {\it No}
non-parametric or parametric quantitative classification system is
perfect. Any classification system will suffer from biases inherent to
its basic definition. However, provided a given quantitative system is
clearly defined before its use, its results will be readily
reproducible in their successes {\it and} failure by other
investigators.

The exact shape of bulge profiles remains under debate
\citep[e.g.,][and references therein]{balcells03}. Locally, there
is evidence that the bulges of late-type spiral galaxies may be better
fit by an $n$ = 1 profile, whereas bright ellipticals and the
bulges of early-type spiral galaxies follow an $n$ = 4 profile
\citep{dejong96,courteau96,andredakis98}.  Local late-type galaxies
with $n$ = 1 bulges have $B/T \leq 0.1$ \citep{dejong96}.  Since such
bulges contain only 10\% of the total galaxy light, low
signal-to-noise measurements of late-type high-redshift galaxies make
it very difficult, if not impossible, to determine the S\'ersic index
of distant bulges even with the spatial resolution of the Hubble Space
Telescope as demonstrated by an extensive set of tests on HST images
of the high-redshift cluster CL1358+62 \citep{tran03}. On the other
hand, $n$ is more important for bulge-dominated galaxies, and $n$ = 4
is the expected value based on local early-type galaxies.  Knowing
that bright ellipticals and the bulges of early-type spirals are
well-fit by a de Vaucouleurs profile, a $n$ = 4 bulge profile was
therefore adopted as the canonical bulge fitting model here for the
sake of continuity across the full range of morphological types.

\subsubsection{Fitting Regions}\label{fitregions}
GIM2D disk+bulge decompositions are performed on thumbnail (or
``postage stamp'') images extracted around the objects detected by
SExtractor rather than on the entire science image itself. The area of
the thumbnail images is given by the isophotal area of the object.
Here, all thumbnails were chosen to have an area 5 times larger than
the 1.5$\sigma_{bkg}$ isophotal area.  Each thumbnail is a square
image with sides of length $\sqrt{5 \times isophotal\_area}$. The first
thumbnail is extracted from the science image itself, and the local
background calculated by SExtractor is subtracted from it so that it
should have a background mean level close to zero.  The second
thumbnail is extracted from the SExtractor segmentation image. The
GIM2D decompositions were performed on all pixels flagged as object
{\it or} background in the SExtractor segmentation image.  Object
areas in the segmentation image are sharply delineated by the location
of the isophote corresponding to the detection threshold because
SExtractor considers all pixels below this threshold to be background
pixels. However, precious information on the outer parts of the galaxy
profile may be contained in the pixels below that threshold, and fits
should therefore not be restricted only to object pixels to avoid
throwing that information away. Pixels belonging to objects in the
neighborhood of the primary object being fit are masked out of the
fitting area using the SExtractor segmentation image. The flux from
the primary object that would have been in those masked areas in the
absence of neighbors is nonetheless properly included in the magnitude
measurements given in this paper because magnitudes were obtained by
integrating the best-fit models over {\it all} pixels.

\subsubsection{Sky Background Level Measurements}\label{skyestm}
Special care must be paid to the determination of the local sky
background level $b$ and dispersion $\sigma_{bkg}$ as sky errors are
the dominant source of systematic errors in bulge+disk decompositions
of distant galaxies. As an example, overestimating the background sky
level will lead to underestimates of the galaxy total flux, half-light
radius and bulge fraction as a result of strong parameter
covariances. Even though the SExtractor local background was
subtracted from each galaxy thumbnail image, an additional (residual)
background estimate $db$ was computed and used by GIM2D to correct for
any systematic error in the initial SExtractor sky level estimate. In order to
compute $db$, GIM2D used all the pixels in the science thumbnail image
flagged as background pixels (flag value of zero) in the SExtractor
segmentation image.  GIM2D further pruned this sample of background
pixels by excluding any background pixel that is closer than five
pixels ($1\farcs0$ for the pixel sampling of the FORS2 detectors)
from any (primary or neighboring) object pixels. This buffer zone
ensures that the flux from all SExtracted objects in the image below
all the 1.5$\sigma_{bkg}$ isophotes does not significantly bias the
mean background level upwards and artificially inflate
$\sigma_{bkg}$. A minimum of 7500 sky pixels was imposed on the area
of the sky region. In cases where the number of sky pixels in the
input science thumbnail image was insufficient, the original
science image was searched for the 7500 sky pixels nearest to the
object.  For the EDisCS fits, background parameters
were re-calculated with GIM2D before fitting, and the residual
background levels $db$ were then frozen to their recalculated values
for the bulge+disk fits. 

\subsubsection{Point-Spread-Functions}
The shape of the Point-Spread-Function (PSF) on the VLT/FORS2 and
HST/ACS images varies significantly as a function of position, and
these variations must be taken into account when
point-spread-functions for the bulge+disk decompositions are
generated. For both sets of images, we used the stand-alone version of
the stellar photometry program DAOPHOT II \citep{stetson87} to
construct spatially-varying PSF models for the EDisCS cluster
images. For each cluster and for each passband, we selected ``clean'',
point sources (detection flag of zero and stellarity index of 0.8 or
greater) from the SExtractor source catalog. The positions of these
point sources were fed to the DAOPHOT routine PSF to be modelled as
the sum of a Gaussian core and of an empirical look-up table
representing corrections from the best-fitting Gaussian to the actual
observed values. Both the gaussian core parameters and the look-up
table were allowed to vary linearly as a function of $x$ and $y$
positions on the image. Finally, the PSF model was used to create a
PSF at the position of each galaxy to be fit. The PSF images were
$2\farcs5$ on a side to provide good dynamical range for the fits.

\subsubsection{Reliability Tests}\label{reliability}
Following the same procedure as in \citet{simard02}, we performed an
extensive set of simulations to test the reliability of our sky
background estimates and of the best-fit parameter values recovered
through bulge+disk fits on both sets of images. 2000 smooth galaxy
image models were created with structural parameters uniformly
generated at random in the following ranges: $20.0 \leq I \leq 25.0$,
$0.0 \leq B/T \leq 1.0$, $0 \leq r_{e} \leq 10\farcs 0$, $0.0 \leq e
\leq 0.7$, $0 \leq r_{d} \leq 10\farcs0$, and $0 \degr \leq i \leq 85 \degr$.
The bulge S\'ersic index was held fixed at $n = 4$ for all models.  Both
bulge and disk position angles were fixed to
90$\degr$\thinspace\thinspace for all simulations, and the bulge and
disk sizes were uniformly generated in the log of the size ranges
above.  Each simulation was convolved with a PSF computed from one of
the images with a FWHM typical of the VLT/FORS2 ($\sim 0\farcs 8$) and
HST/ACS ($\sim 0\farcs 05$) observations.  The same PSF was used in
both creating and analyzing the simulations, so the results will not
include any error in the structural parameters due to PSF mismatch.
Poisson deviates were used to add photon noise due to galaxy flux into
the simulations.  The noisy images were then embedded in a 20$\arcsec$
$\times$ 20\arcsec section of one of the real $I-$band images to
provide a real background for the simulations.  In addition to sky
photon noise and detector read-out noise, the real background noise
includes brightness fluctuations of very faint galaxies below the
detection threshold.  This procedure thus yields realistic errors that
include the effect of sky errors. The simulations were SExtracted
exactly in the same way as real EDisCS sources (see
Section~\ref{sources}). Science and segmentation thumbnails extracted
from the simulations were analyzed with GIM2D following exactly the
same steps as for the real galaxies (see Section~\ref{bdcomps}).

\begin{figure*}
\resizebox{\hsize}{!}{\includegraphics[angle=270]{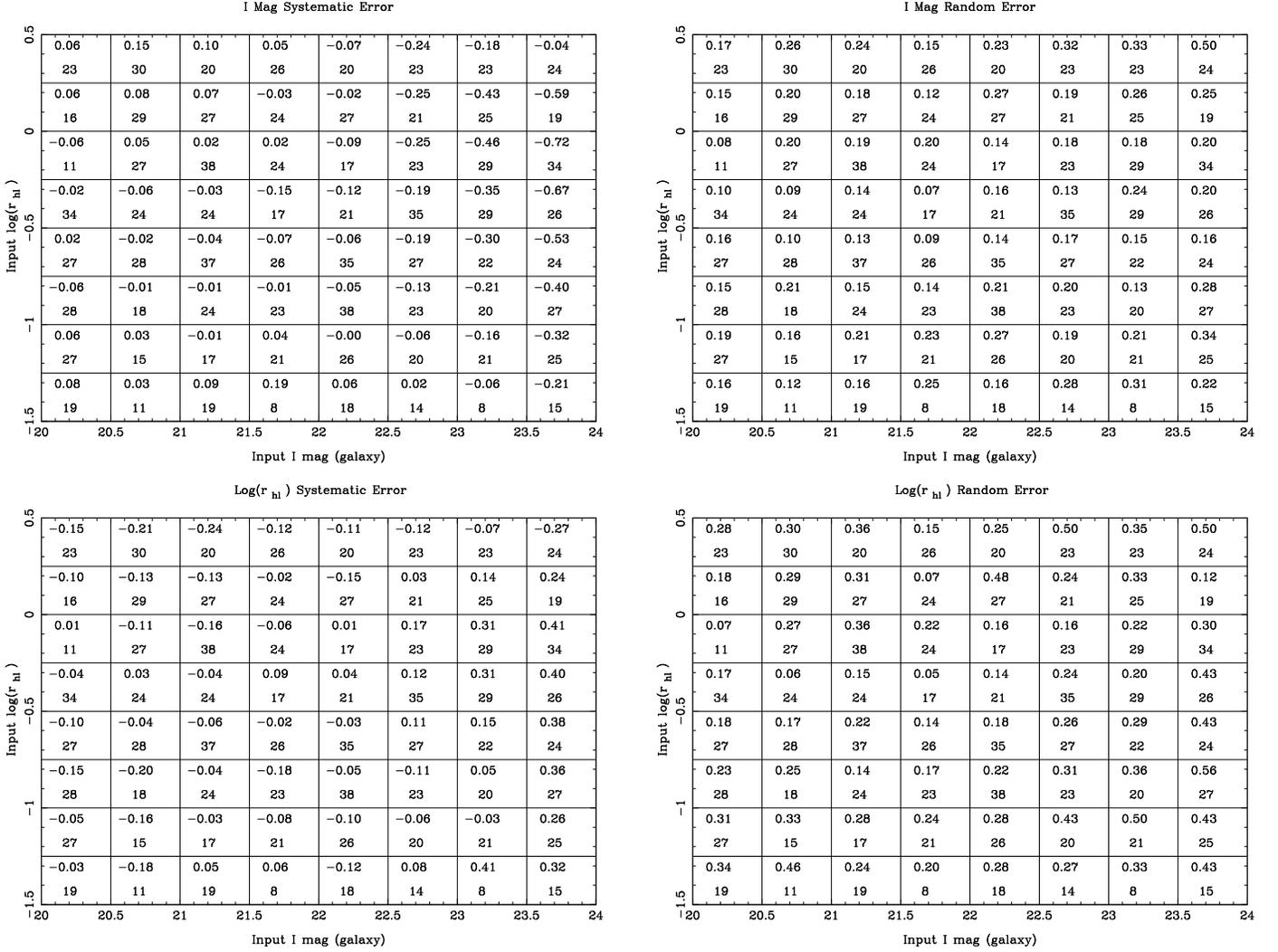}}
\caption{Two-dimensional maps of GIM2D systematic and random galaxy
magnitude and half-light radius errors from 2000 VLT/FORS2 image
simulations. {\it Top left-hand panel}: Systematic error on recovered
galaxy total magnitude $I_{rec}$ as a function of {\it input} galaxy
log half-light radius $r_{hl,input}$ in arcseconds and {\it input}
galaxy total magnitude $I_{input}$.  The top number in each cell is
the mean magnitude error ($I_{rec} - I_{input}$), and the bottom
number is the number of simulations created in that cell. {\it Top
right-hand panel}: 1$\sigma$ random error on $I_{rec}$
($\sigma$($I_{rec}-I_{input}$)) as a function of log $r_{hl,input}$
and $I_{input}$. {\it Bottom left-hand panel}: Systematic error on
recovered galaxy intrinsic log half-light radius $r_{hl,rec}$ as a
function of {\it input} galaxy log half-light radius $r_{hl,input}$ in
arcseconds and {\it input} galaxy total magnitude $I_{input}$.  The
top number in each cell is the mean log radius error (log $r_{hl,rec}
-$ log $r_{hl,input}$), and the bottom number is the number of
simulations created in that cell. {\it Top right-hand panel}:
1$\sigma$ random error on log $r_{hl,rec}$ ($\sigma$(log
$r_{hl,rec}-$ log $r_{hl,input}$)) as a function of log $r_{hl,input}$
and $I_{input}$.}
\label{emap_m-rhl}
\end{figure*}

\begin{figure*}
\begin{center}
\resizebox{\hsize}{!}{\includegraphics[angle=270]{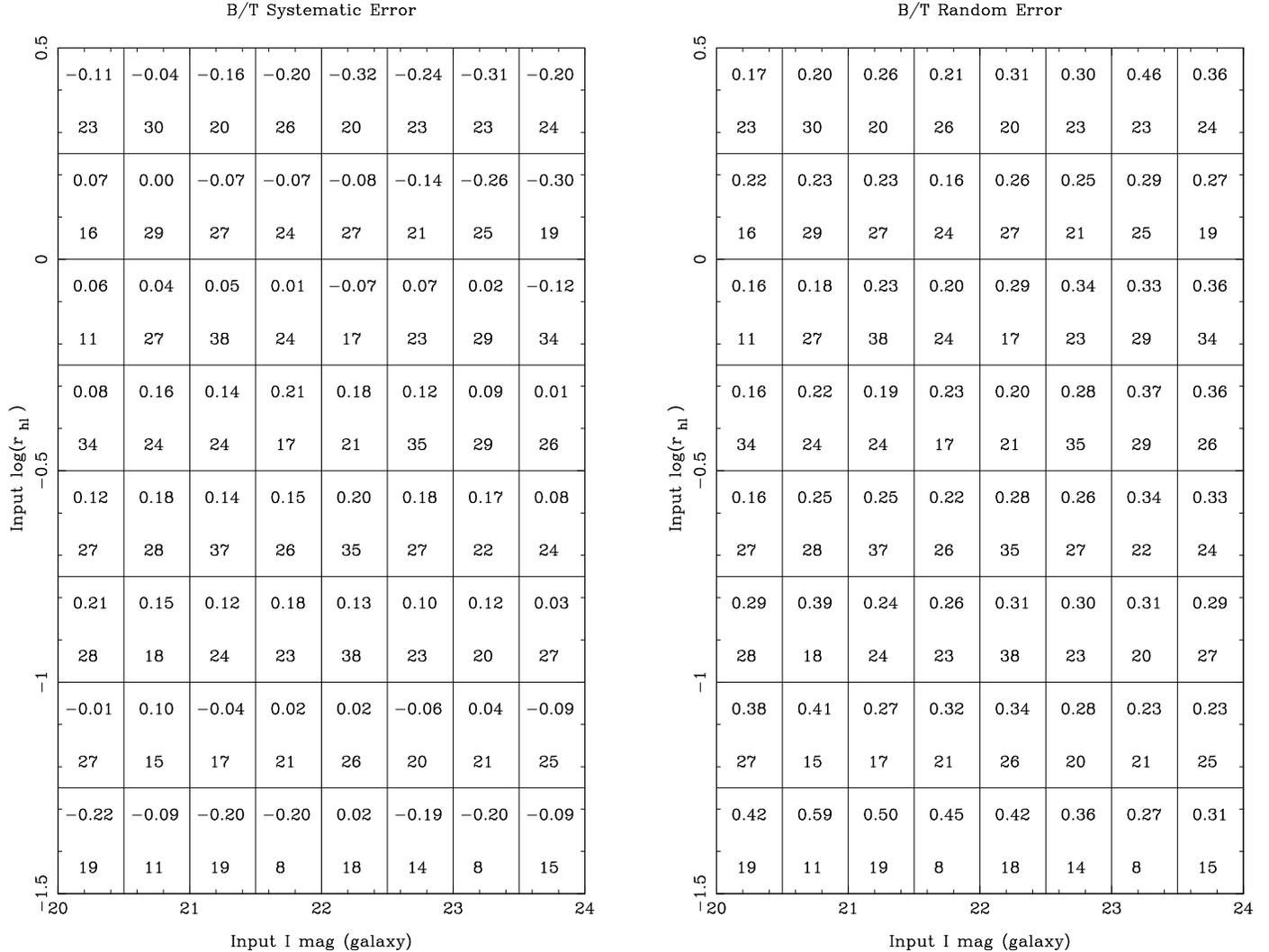}}
\caption{Two-dimensional maps of GIM2D systematic and random galaxy
bulge fraction errors from 2000 VLT/FORS2 image simulations. {\it Top
left-hand panel}: Systematic error on recovered galaxy bulge fraction
$(B/T)_{rec}$ as a function of {\it input} galaxy log half-light
radius $r_{hl,input}$ in arcseconds and {\it input} galaxy total
magnitude $I_{input}$.  The top number in each cell is the mean bulge
fraction error ($(B/T)_{rec} - (B/T)_{input}$), and the bottom number
is the number of simulations created in that cell. {\it Top right-hand
panel}: 1$\sigma$ random error on $(B/T)_{rec}$
($\sigma$($(B/T)_{rec}-(B/T)_{input}$)) as a function of log
$r_{hl,input}$ and $I_{input}$.}
\label{emap-bt}
\end{center}
\end{figure*}

Figures~\ref{emap_m-rhl} and~\ref{emap-bt} show maps of errors on the
galaxy total magnitude $I$, galaxy intrinsic half-light radius $r_{hl}$ and
galaxy bulge fraction $B/T$ for the VLT/FORS2 images. The left-hand
panels show the mean parameter errors as a function of input
galaxy magnitude and size, and the right-hand panels show the
1$\sigma$ parameter random error as a function of input galaxy
magnitude and size. The lower number in each cell is the number of
simulated galaxies created for that cell. Most systematic errors are
directly related to surface brightness as magnitudes and sizes of low
surface brightness sources are inherently harder to measure. This fact
is borne out by the trends in the errors shown in
Figure~\ref{emap_m-rhl}. Decreasing surface brightness follows a line
going from the lower left-hand corners to the upper right-hand
ones. The top panels of Figures~\ref{emap_m-rhl} show that systematic
errors on $I$ start to become significant ($\Delta I \simeq 0.2$)
fainter than $I$ = 22.5. Systematic errors on log $r_{hl}$ also
increases significantly beyond this magnitude.  It is important to
note that $I$ = 22.5 is significantly fainter by about 2 mag than the
galaxies that will be used to compute cluster early-type galaxy
fractions in Section~\ref{efrac-VLT}, so these galaxy fractions should be
unaffected. Figure~\ref{emap-bt} shows that systematic errors on $B/T$
are smallest over the region $I \leq 22.5, -0.5 \leq $ log $r_{hl}
\leq 0.3$ where most of the real EDisCS galaxies actually lie. As mentioned above, our reliability tests do not include the effects of PSF mismatch errors because we used the same PSF for creating simulated images and for their analysis. However, we were able to check that these errors were not significant because we fitted both galaxies {\it and} stars on our real VLT/FORS2 images. The measured intrinsic radii of the stars clustered at zero, and this would not have been the case should PSF mismatch errors have been important.

\section{Early-Type Galaxy Fractions}\label{efractions}

\subsection{Definition and Comparison with Galaxy Visual Classifications}\label{efrac-defn}

The bulk of the previous work on the morphological content of
high-redshift clusters is based on the visual classification of
galaxies, and this section compares visual and quantitative
morphological classification. Visual classifications for 9200 galaxies
in EDisCS clusters with HST images are presented in \citet{desai07}.
As shown by previous works \citep{im02,mcintosh02,tran03,blakeslee06},
quantitative and visual morphologies can be best linked together by
focussing on three structural parameters: bulge fraction $B/T$, image
smoothness $S$ and bulge ellipticity $e$. The image smoothness, $S$,
is defined as:

\begin{equation}
S = R_T + R_A
\label{image-smoothness}
\end{equation}

\noindent where $R_T$ and $R_A$ are defined in Equation 11 of \citet{simard02}. These two indices quantify the amount of light in symmetric and asymmetric residuals from the fitting model respectively, and they are expressed as a fraction of the total galaxy model flux. $S$ is typically measured inside a radius that is a
multiple of the galaxy half-light radius. Using our HST/ACS
measurements, we found no differences between image smoothness within
one and two galaxy half-light radii. We therefore use image
smoothness inside two half-light radii (and denote it $S2$ hereafter)
because it is more reliably measured on the VLT/FORS2 images with
their lower spatial resolution. We can choose selection criteria on
$B/T$, $S$ and $e$ that yield the best match to the visual
classifications, and the particular choices are not important as long
as the same selection criteria are applied to both local and
high-redshift clusters.
 
\begin{figure}
\resizebox{\hsize}{!}{\includegraphics[angle=270]{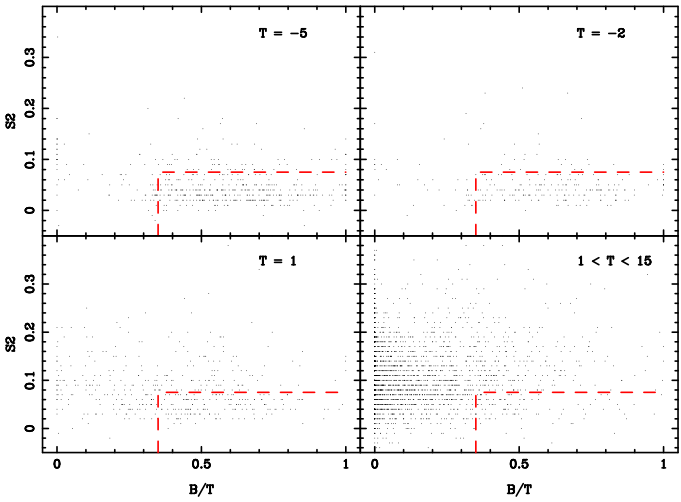}}
\caption{Image smoothness parameter $S2$ versus bulge fraction $B/T$ for different visual types. The galaxies selected by our quantitative early-type galaxy criteria ($B/T \geq 0.35$ and $S2 \leq 0.075$) are enclosed in the area delimited by dashed lines.}  
\label{hst_visual_S2-vs-bt}
\end{figure}

We divide the visually-classified EDisCS into $T$ = $-$5 (E), $-$2 (S0), 1
(S0/a) and ``others'' ($T > 1$). Using our HST/ACS structural
parameter measurements, we find that E and S0 galaxies have similar
$B/T$ distribution with the S0 distribution being skewed towards
slightly lower $B/T$, but $e$ distributions are different. It is
therefore possible to differentiate between E and S0
galaxies on the basis of these two parameters. S0 and S0/a galaxies
have similar $e$ distributions but different $B/T$ and $S$
distributions. Given that the bulge ellipticity $e$ cannot be reliably
measured on the VLT/FORS2 images, we restrict on selection criteria to
$B/T$ and $S2$. Figure~\ref{hst_visual_S2-vs-bt} shows $S2$ versus
$B/T$ for the four visual types of galaxies. $S2$ can take on small
negative values due to statistical background subtraction terms
\citep{simard02}. The optimal choice of limits on $B/T$ and $S2$ for
our definition of early-type fraction is driven by the
need to maximize the number of E/S0 galaxies selected while minimizing
the contamination from Sa-Irr galaxies. After several iterations, we
settled on $B/T \geq 0.35$ and $S2 \leq 0.075$ as our definition of an early-type galaxy. These limits are very similar to
those used in previous studies \citep{im02,mcintosh02,tran03}. With
these criteria, our quantitative selection can be translated
into visual classification terms as

\begin{equation}
f_{et} = \bigr(0.69 N_E + 0.71 N_{S0} + 0.35 N_{S0/a} + 0.04 N_{Sa-Irr}\bigl) / N_{total}
\label{fet-gim2d}
\end{equation}

\noindent The coefficients in Equation~\ref{fet-gim2d} give the completeness of the quantitative classification in terms of the \citet{desai07} visual classes. For example, the adopted $B/T$ and $S2$ cuts would select 69$\%$ of the galaxies visually classified by \cite{desai07} as E's, 71$\%$ of their S0's and so on. As mentioned earlier, E's and S0's cannot be distinguished using only $B/T$ and $S2$. Equation~\ref{fet-gim2d} is to be compared to the prescription of \citet{dokkum00}:
\begin{equation}
f_{et} = \bigr(N_E + N_{E/S0} + N_{S0} + \frac{1}{2} N_{S0/a}\bigl) / N_{total}
\label{fet-vdk}
\end{equation}
\noindent where $N_{total}$ is the number of galaxies with $M_{V} \leq -20$. 

It is impossible to recover all the galaxies visually classified as
early-types because a visual early-type does not necessarily imply a
$r^{1/4}$ profile. Indeed, many early-type galaxies such as dwarf
ellipticals have simple exponential profiles \citep{lin83,kormendy85},
and we have verified through isophote tracing that many galaxies
visually classified as early-types and missed by our selection
criteria do have radial surface brightness profiles that are
exponential and thus consistent with their measured low $B/T$ values.

Given $N_{total}$ galaxies brighter than an absolute magnitude limit
$M_{V,lim}$ inside a clustercentric radius $R_{max}$ of which $N_{et}$
are early-types galaxies, we actually calculate the early-type galaxy
fraction by finding the median of the binomial probability
distribution

\begin{equation}
p(x) dx = {{N_{total}!}\over{N_{et}!(N_{total}-N_{et})!}} x^{N_{et}}(1-x)^{N_{total}-N_{et}}
\label{binom_dist}
\end{equation}
\noindent and we integrate Equation~\ref{binom_dist} to calculate the lower and upper bounds of the corresponding 68$\%$ confidence interval. In the limit of large $N_{total}$ and $N_{et}$ (not always true for the current cluster sample), this converges to the same symmetric error bars as would be obtained from the propagation of gaussian errors.

\subsection{HST-Based Fractions}\label{efrac-HST}

For each EDisCS cluster with HST/ACS imaging, we have computed the
fraction of early-type galaxies using our quantitative
HST/ACS morphologies ($B/T \geq 0.35$ and and $S2 \leq 0.075$). We
used only spectroscopically-confirmed members brighter than an
absolute $V$-band magnitude $M_{V,lim}$. We varied $M_{V,lim}$ as a
function of redshift from $-$20.5 at $z$ = 0.8 to $-$20.1 at $z$ = 0.4
to account for passive evolution. This choice of $M_{V,lim}$ was made to be fully consistent with previous work \citep{poggianti06} although it may not be strictly the best choice for late-type galaxy populations. Our results did not appear to be sensitive to variations in $M_{V,lim}$ at the level of a few tens of a magnitude. Following \citet{poggianti06}, our early-type galaxy fractions were also computed by weighting each galaxy
according to the incompleteness of the spectroscopic catalog. This
incompleteness depends on both galaxy magnitude and clustercentric
position. Incompleteness as a function of magnitude was computed by
dividing the number of galaxies in the spectroscopic catalog in a
given magnitude bin by the number of galaxies in the parent
photometric catalog in the same bin. We used 0.5 mag bins here.
Incompleteness due to the geometrical effects comes from the finite
number of slitlets per sky area, and the increasing surface density of
galaxies on the sky closer to the cluster centers. Geometric
incompleteness is field dependent as it depends on cluster richness,
and we thus computed this incompleteness on a field-by-field basis. We
also used four radial bins out to $R_{200}$ with a bin width of
0.25$R_{200}$.

The raw and incompleteness-corrected HST-based early-type galaxy
fractions are given in Table~\ref{HST-efracs-table} for a maximum
clustercentric radius $R_{et}$ of 0.6$R_{200}$ (columns 4 and 5) and
$R_{200}$ (columns 9 and 10). Most of the corrected fractions do not
significantly differ from the raw ones because our spectroscopic
sample is essentially complete down to $I \leq 23$ ($M_{V} \sim $ $-$20 at
$z = 0.8$), and we used multiple masks on dense clusters to improve
the spatial sampling of our spectroscopic sample. As a comparison,
Table~\ref{HST-efracs-table} also gives early-type galaxy fractions
measured from visual classifications by \citet{desai07} (Columns 6 and
7). They should be compared with values in Column 5 because cluster galaxy samples selected using photometric redshifts are {\it de facto} free from the magnitude and geometric incompleteness of our spectroscopic sample.  Another
important caveat is that they were computed using two different ways to
isolate cluster members (photometric redshift and statistical
background subtraction), and they are thus not restricted to
spectroscopically-confirmed members. Nonetheless, the agreement
between fractions measured from visual and quantitative
classifications is remarkably good.  The largest disagreement is for 1138.2-1133, but even this case can be considered marginal as it is not quite 2$\sigma$. 
\begin{table*}
\caption[]{Early-Type Galaxy Fractions Based on HST/ACS Imaging}
\begin{center}
\begin{tabular*}{16cm}{ccccccccccc}
\hline
ID & \multicolumn{6}{c}{$R_{et} \leq 0.6 R_{200}$} & \multicolumn{4}{c}{$R_{et} \leq R_{200}$}\\
 & $N_{clus}$ & $N_{et}$ & $f_{et,raw}$ & $f_{et,corr}$ & $f_{E/S0,phz}^{\rm a}$ & $f_{E/S0,bkg}^{\rm b}$ & $N_{clus}$ & $N_{et}$ & $f_{et,raw}$ & $f_{et,corr}$\\
(1) & (2) & (3) & (4) & (5) & (6) & (7) & (8) & (9) & (10) & (11)\\
\hline
1216.8-1201 & 45 & 23 & 0.51$\pm$0.07 & 0.55$\pm$0.07 & 0.47$\pm$0.06 & 0.54$\pm$0.06 & 57 & 25 & 0.44$\pm$0.06 & 0.42$\pm$0.06 \\
1040.7-1156 & 9 & 4 & 0.45$\pm$0.15 & 0.45$\pm$0.15 & 0.53$\pm$0.17 & 0.37$\pm$0.17 & 13 & 5 & 0.40$\pm$0.12 & 0.33$\pm$0.12\\
1054.4-1146 & 18 & 9 & 0.50$\pm$0.11 & 0.29$\pm$0.10 & 0.28$\pm$0.09 & 0.24$\pm$0.09 & 26 & 11 & 0.43$\pm$0.10 & 0.28$\pm$0.09\\
1054.7-1245 & 11 & 8 & 0.70$\pm$0.13 & 0.46$\pm$0.14 & 0.44$\pm$0.16 & 0.57$\pm$0.13 & 19 & 10 & 0.52$\pm$0.11 & 0.38$\pm$0.11\\
1232.5-1250 & 48 & 28 & 0.58$\pm$0.07 & 0.58$\pm$0.07 & 0.60$\pm$0.06 & 0.45$\pm$0.06 & 51 & 28 & 0.55$\pm$0.07 & 0.47$\pm$0.07\\
1037.9-1243 & 8 & 1 & 0.18$\pm$0.12 & 0.18$\pm$0.12 & 0.13$\pm$0.09 & 0.08$\pm$0.07 & 8 & 1 & 0.18$\pm$0.12 & 0.18$\pm$0.12\\
1103.7-1245b & 2 & 0 & 0.21$\pm$0.20 & 0.21$\pm$0.20 & 0.47$\pm$0.20 & 0.21$\pm$0.20 & 4 & 0 & 0.13$\pm$0.13 & 0.13$\pm$0.13 \\
1354.2-1231 & 8 & 5 & 0.61$\pm$0.15 & 0.39$\pm$0.15 & 0.35$\pm$0.14 & 0.44$\pm$0.19 & 12  & 5 &  0.42$\pm$0.13 & 0.28$\pm$0.11\\
1138.2-1133 & 22 & 4 & 0.20$\pm$0.08 & 0.20$\pm$0.08 & 0.37$\pm$0.09 & 0.50$\pm$0.14 & 24 & 6 & 0.26$\pm$0.08 & 0.34$\pm$0.09\\
\hline
\end{tabular*}
\label{HST-efracs-table}

$^{\rm a}$ From Table 14 of \citet{desai07}\\
$^{\rm b}$ From Table 16 of \citet{desai07}
\end{center}
\end{table*}
 
\subsection{VLT- versus HST-based Fractions}\label{efrac-VLT}

\begin{figure*}
\resizebox{\hsize}{!}{\includegraphics[angle=270]{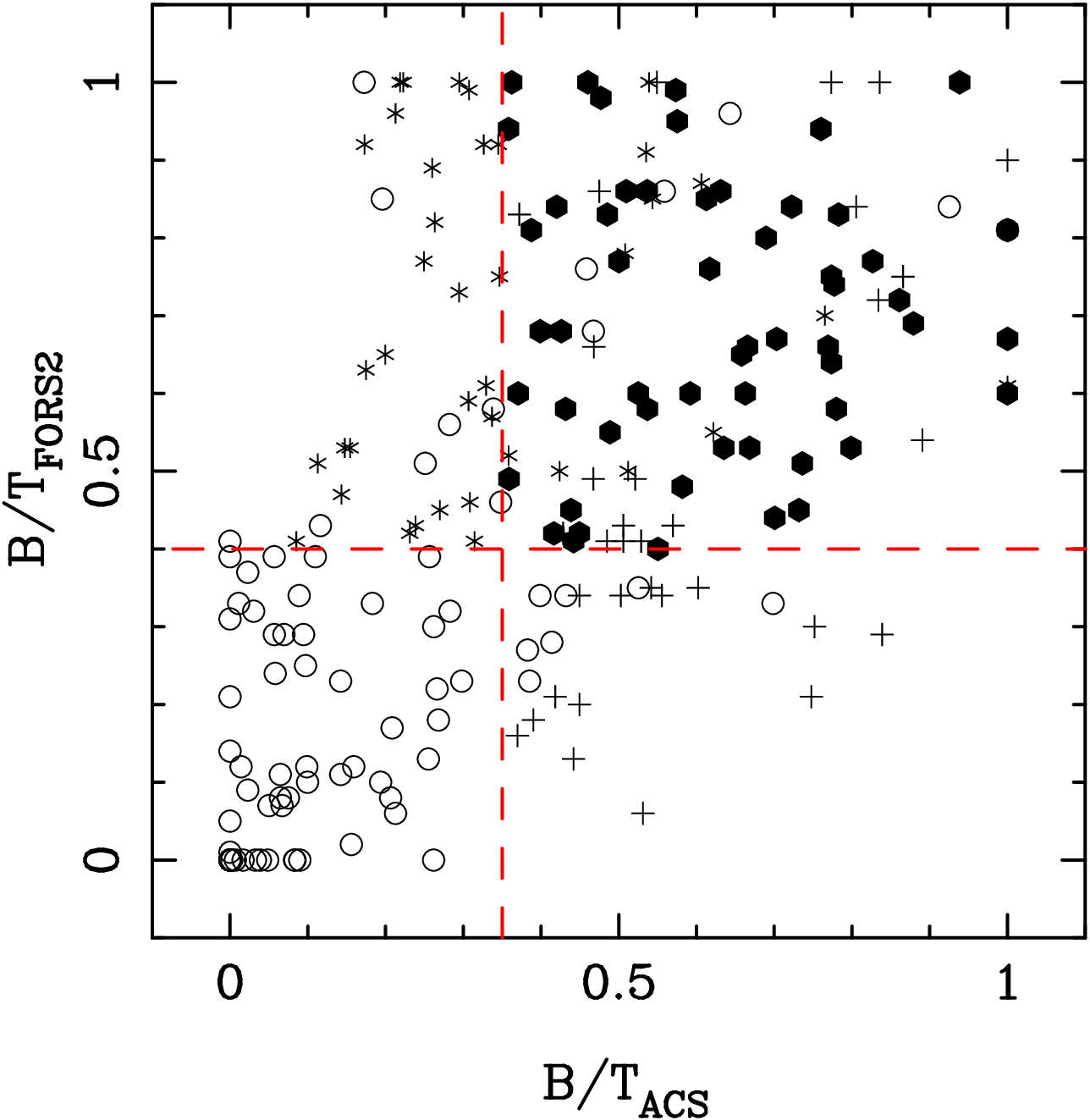},\includegraphics[angle=270]{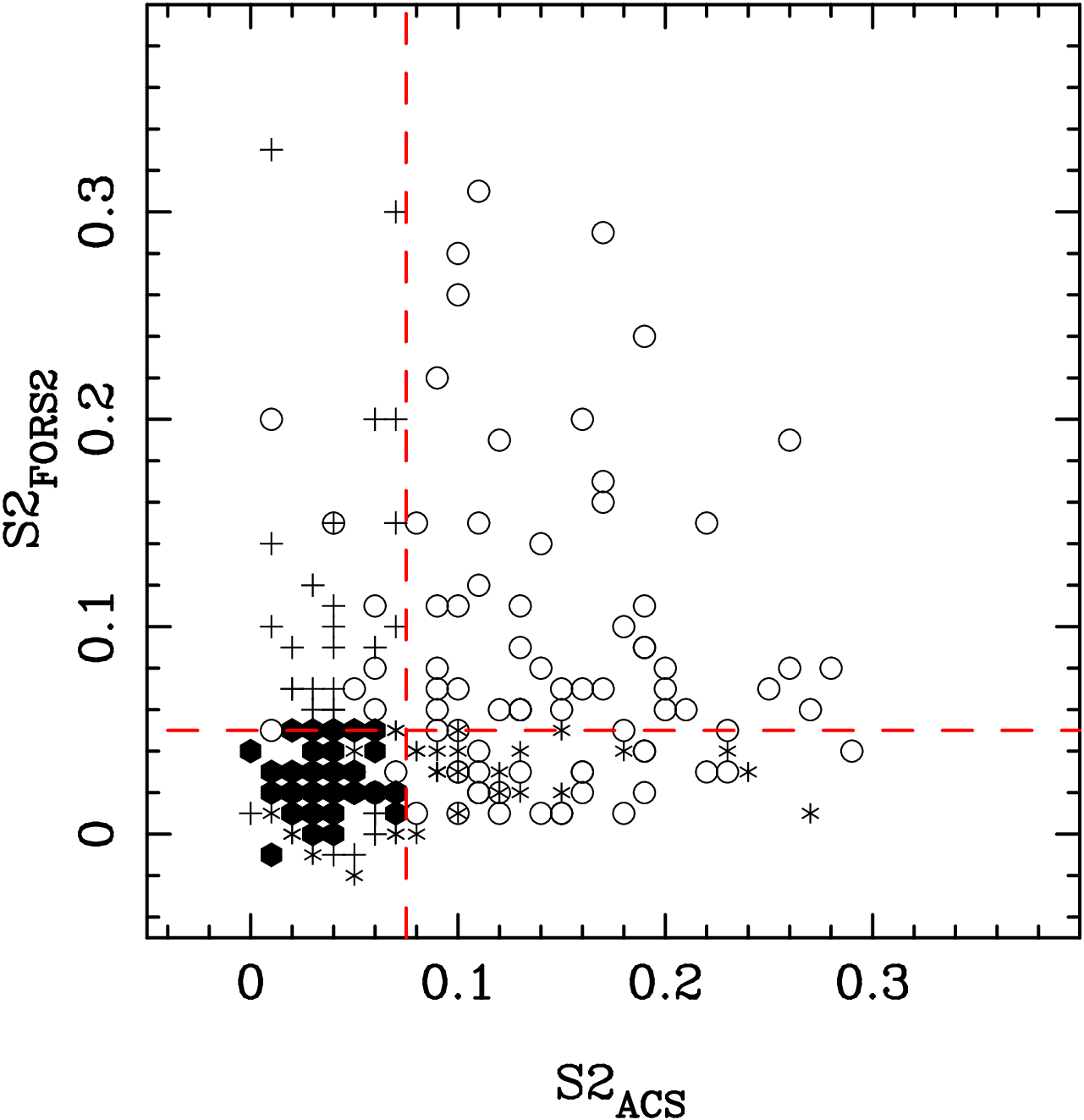}}
\caption{Direct galaxy-by-galaxy comparison between bulge fraction (left-hand panel) and image smoothness (right-hand panel) measurements from HST/ACS and VLT/FORS2 images. Filled circles are galaxies classified as early-type on both ACS and VLT images, asterisks are galaxies classified as early-type only on the VLT images, pluses are galaxies classified as early-type only on the ACS images, and open circles are galaxies not classified as early-type on either ACS or VLT images, The dashed lines show the cuts used for the definition of an early-type galaxy as discussed in Sections~\ref{efrac-defn} and~\ref{efrac-VLT}.}
\label{hst-vlt_morph}
\end{figure*}

Quantitative morphologies measured from HST images are more robust
than those measured from ground-based images
(Section~\ref{reliability} and
\citet{simard02}). Figure~\ref{hst-vlt_morph} shows a direct
galaxy-by-galaxy comparison between bulge fraction and image
smoothness measurements from HST/ACS and VLT/FORS2 images. This
comparison includes spectroscopically-confirmed member galaxies from
all clusters with HST imaging that are brighter than $M_{V,lim}$ and
within a clustercentric radius of 0.6$R_{200}$ to take into account
the effect of crowding. For a given galaxy, the agreement between the
two sets of measurements will obviously depend on its apparent
luminosity and size. The overall agreement is reasonably good. The
scatter in the bulge fraction plot is consistent with $\sigma_{B/T,
ACS}$ $\sim$ 0.1 \citep{simard02} and $\sigma_{B/T, VLT} \sim$ 0.25
(Figure~\ref{emap-bt}) added in quadrature, but the fact that
completely independent segmentation images were used for the HST and
VLT morphological measurements also contributes significantly to this
scatter. Indeed, this scatter would be smaller if only uncrowded
galaxies (as indicated by the SExtractor photometry flag) on the VLT
images had been plotted here. For the image smoothness plot, there is
a correlation between $S2_{FORS2}$ and $S2_{ACS}$, but it is not
one-to-one. $S2_{ACS}$ values increase faster than $S2_{FORS2}$. This
is expected as PSF blurring will be more significant on the
ground-based images, and $S2$ measurements are not corrected for PSF
effects. Part of the scatter is again due to the use of independent
segmentation images.

\begin{figure}
\resizebox{\hsize}{!}{\includegraphics[angle=270]{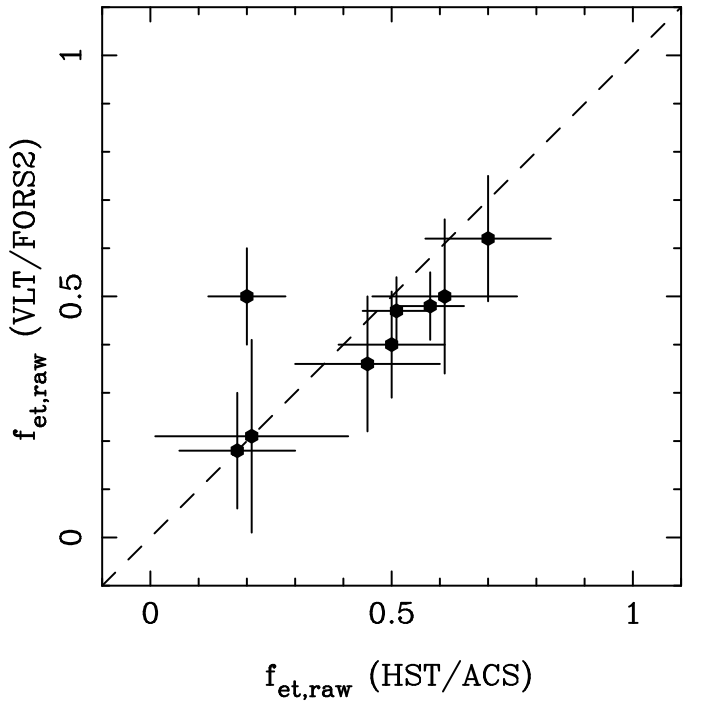}}
\caption{Comparison between early-type galaxy fractions for clusters with overlapping VLT and HST imaging. VLT/FORS2 and HST/ACS early-type galaxy fractions were computed using galaxies with $B/T_{FORS2} \geq 0.40$ and $S2_{FORS2} \leq 0.05$ and $B/T_{ACS} \geq 0.35$ and $S2_{ACS} \leq 0.075$ respectively. The ACS and FORS2 $f_{et}$ values plotted here are listed in column 4 of Table~\ref{HST-efracs-table} and column 4 of Table~\ref{VLT-efracs-table}. Dashed line is the one-to-one line.}
\label{hst-vlt_efrac}
\end{figure}

The inclusion of clusters with only VLT/FORS2 imaging allows us
to extend our analysis to nine additional clusters - an important
consideration given that we seek to probe cluster-to-cluster
variations in morphological content. We therefore need to show that we
measure consistent early-type fractions for clusters with overlapping
ACS and FORS2 images. The problem boils down to finding the set of
limits on $B/T_{FORS2}$ and $S2_{FORS2}$ that yield FORS2 early-type
fractions in agreement with the ACS fractions obtained with $B/T_{ACS}
\geq 0.35$ and $S2_{ACS} \leq 0.075$ when the same galaxies are used
for both FORS2 and ACS. For each cluster, we used all
spectroscopically-confirmed cluster members brighter than $M_{V,lim}$
and within a clustercentric radius of $R_{200}$. No corrections for
incompleteness were applied here as these corrections would be
identical for both cases.  We went through many manual iterations
until we found satisfactory limits on $B/T_{FORS2}$ and
$S2_{FORS2}$. We found FORS2 fractions to be in very good agreement
with the ACS ones for $B/T_{FORS2} \geq 0.40$ and $S2_{FORS2} \leq
0.05$ (Figure~\ref{hst-vlt_efrac}). This agreement is especially good
if one considers the fact that we performed our FORS2 and ACS
bulge+disk decompositions completely independently from one another,
i.e., we did not attempt to use the same SExtractor segmentation map
for both FORS2 and HST images. The limit on $B/T_{FORS2}$ is slightly
higher than the one on $B/T_{ACS}$ because lower spatial resolution
typically leads to a small overestimate of the bulge
fraction. Similarly, the limit on $S2_{FORS2}$ needs to be more
stringent than on $S2_{ACS}$ to select the same galaxies as they will
look smoother on the FORS2 images due to lower resolution.

Following the procedure described in Section~\ref{efrac-HST}, we
computed early-type galaxy fraction for all eighteen clusters using
galaxies on our FORS2 images with $B/T_{FORS2} \geq 0.40$ and
$S2_{FORS2} \leq 0.05$. The results are shown in
Table~\ref{VLT-efracs-table}. The same incompleteness corrections as in
Section~\ref{efrac-HST} were applied here as well. The errors on the
early-type galaxy fractions in the table do not include errors
on $R_{200}$ due to correlated errors on cluster $\sigma$. We hereafter use our VLT/FORS2 early-type fractions for all EDisCS clusters for the sake of uniformity.

\begin{table*}
\caption[]{Early-Type Galaxy Fractions Based on VLT/FORS2 Imaging}
\begin{center}
\begin{tabular*}{12cm}{cccccccc}
\hline
ID & $M_{V,lim}$ & \multicolumn{3}{c}{$R_{et} \leq 0.6 R_{200}$} & \multicolumn{3}{c}{$R_{et} \leq R_{200}$}\\
 & & $N$$^{\rm a}$ & $f_{et,raw}$ & $f_{et,corr}$ & $N$$^{\rm a}$ & $f_{et,raw}$ & $f_{et,corr}$ \\
(1) & (2) & (3) & (4) & (5) & (6) & (7) & (8) \\
\hline
1018.8-1211 & $-$20.2 & 18 & 0.60$\pm$0.11 & 0.55$\pm$0.11 & 20 & 0.59$\pm$0.11 & 0.55$\pm$0.11 \\
1059.1-1253 & $-$20.2 & 20 & 0.64$\pm$0.10 & 0.59$\pm$0.10 & 28 & 0.64$\pm$0.09 & 0.60$\pm$0.09 \\
1119.3-1130 & $-$20.3 & 6 & 0.64$\pm$0.17 & 0.77$\pm$0.15 & 9 & 0.64$\pm$0.14 & 0.74$\pm$0.15 \\
1202.7-1224 & $-$20.1 & 11 & 0.54$\pm$0.14 & 0.54$\pm$0.14 & 13 & 0.60$\pm$0.13 & 0.67$\pm$0.13 \\
1232.5-1250 & $-$20.2 & 48 & 0.48$\pm$0.07 & 0.60$\pm$0.07 & 51 & 0.49$\pm$0.07 & 0.62$\pm$0.07 \\
1301.7-1139 & $-$20.2 & 17 & 0.53$\pm$0.11 & 0.53$\pm$0.11 & 28 & 0.43$\pm$0.09 & 0.40$\pm$0.09 \\
1353.0-1137 & $-$20.3 & 9 & 0.55$\pm$0.15 & 0.55$\pm$0.15 & 17 & 0.42$\pm$0.11 & 0.36$\pm$0.11 \\
1411.1-1148 & $-$20.2 & 15 & 0.47$\pm$0.12 & 0.53$\pm$0.12 & 16 & 0.44$\pm$0.12 & 0.44$\pm$0.11 \\
1420.3-1236 & $-$20.2 & 4 & 0.50$\pm$0.20 & 0.50$\pm$0.20 & 7 & 0.56$\pm$0.17 & 0.44$\pm$0.17\\
1037.9-1243 & $-$20.3 & 8 & 0.18$\pm$0.12 & 0.18$\pm$0.12 & 8 & 0.18$\pm$0.12 & 0.18$\pm$0.12 \\
1040.7-1156 & $-$20.4 & 9 & 0.36$\pm$0.14 & 0.36$\pm$0.14 & 13 & 0.46$\pm$0.13 & 0.46$\pm$0.13 \\
1054.4-1146 & $-$20.4 & 18 & 0.40$\pm$0.11 & 0.34$\pm$0.11 & 26 & 0.39$\pm$0.09 & 0.35$\pm$0.09 \\
1054.7-1245 & $-$20.5 & 11 & 0.62$\pm$0.13 & 0.38$\pm$0.13 & 19 & 0.52$\pm$0.11 & 0.43$\pm$0.11 \\
1103.7-1245b & $-$20.4 & 2 & 0.21$\pm$0.20 & 0.21$\pm$0.20 & 3 & 0.39$\pm$0.22 & 0.39$\pm$0.22 \\
1138.2-1133 & $-$20.2 & 22 & 0.50$\pm$0.10 & 0.46$\pm$0.10 & 24 & 0.54$\pm$0.10 & 0.54$\pm$0.10 \\
1216.8-1201 & $-$20.5 & 45 & 0.47$\pm$0.07 & 0.53$\pm$0.07 & 57 & 0.46$\pm$0.06 & 0.46$\pm$0.06 \\
1227.9-1138 & $-$20.3 & 9 & 0.26$\pm$0.13 & 0.16$\pm$0.11 & 11 & 0.30$\pm$0.12 & 0.22$\pm$0.11 \\
1354.2-1231 & $-$20.5 & 8 & 0.50$\pm$0.16 & 0.39$\pm$0.15 & 12 & 0.35$\pm$0.12 & 0.28$\pm$0.11 \\
\hline
\end{tabular*}
\label{VLT-efracs-table}

$^{\rm a}$ Number of cluster members brighter than $M_{V,lim}$ inside $R_{et}$
\end{center}
\end{table*}

\subsection{Local Clusters}\label{efrac_sdss}

The Sloan Digital Sky Survey \citep[SDSS;][]{abazajian09} offers by
far the best, ``local'' ($z < 0.1$) baseline for a comparison of
early-type galaxy fractions between local and high-redshift
clusters. Clusters similar in mass to EDisCS clusters can be
selected from spectroscopic SDSS data, {\it and} galaxy morphologies
can be measured using GIM2D from SDSS images. We
therefore used SDSS-selected clusters here to construct a local
baseline as nearly free of systematics as currently possible given the
available data.

We use the sample of SDSS clusters defined in \citet{linden07}.  The basis of this cluster sample is the C4 cluster catalogue \citep{miller05}, and we
briefly recapitulate here how the von der Linden et al. sample was selected. Their primary aim was to find the galaxy closest to the deepest point of the potential well of a cluster. In order to insure that the clusters would span a large angular extent compared to the minimum distance of 55 arcsec between fibers, the sample was restricted to redshifts $z \leq 0.1$. This first cut resulted in an initial sample of 833 clusters. A combination of clustercentric distance, galaxy concentration and colour cuts was used to identify brightest cluster galaxies (BCGs) for these clusters. For cases where the same BCG was identified for more than one cluster, only the cluster with the density peak was retained, and the others were deemed to be substructures. This cut rejected 101 clusters. Refined velocity dispersion and virial radii were then computed through an iterative process of velocity cuts. This process failed for 55 clusters, and these were also rejected. All remaining clusters were then visually inspected. An additional set of 35 clusters were rejected at this point as being in the infall regions of other clusters, and another 17 clusters were discarded because they had less than three galaxies within 3$\sigma$ of the cluster redshift and 1$R_{200}$ of its center. This brought the total of SDSS clusters down to 625. Following \citet{poggianti06}, we applied a final redshift cut to keep clusters in the range 0.04 $< z < $ 0.085. The lower limit reduces fiber aperture effects, and the upper limit minimizes incompleteness in galaxy absolute magnitude. Our final SDSS comparison sample thus has 439 clusters.

Given that we are interested in probing galaxy properties as a function of environment, it is important to ensure that the SDSS and EDisCS samples both cover the same range of environments. We therefore selected a subsample of SDSS clusters with a velocity dispersion distribution matching the EDisCS distribution. This match was done by adding SDSS clusters to the subsample one at a time and keeping only those that maintained the EDisCS-SDSS two-sample Kolmogorov-Smirnov probability above 50$\%$.  This is the probability of the maximum difference between the normalized cumulative distributions of the EDisCS and SDSS samples. It means that even if the two sampls were selected at random from the same underlying distribution, they would differ by more than the two observed samples more than half the time. This probability threshold thus  yields a SDSS subsample that is very well-matched to the EDisCS clusters. The resulting subsample (referred to as ``SDSS-C4" hereafter) includes 158 clusters, and these clusters are listed in Table~\ref{sdss-cls-list}.

We ran GIM2D on SDSS Data Release Seven \citep[DR7;][]{abazajian09} $u$-, $g$-, $r$- and $i$-band images of objects in the
magnitude range $14 \leq r_{petrosian,corr} \leq 17.77$ with a galaxy spectrum
(i.e., with field SpecClass = 2 in database table SpecPhoto). Bulge+disk decompositions were successfully
obtained for 674,693 galaxies (Simard, in
preparation). GIM2D morphologies for galaxies in our matched SDSS-C4 clusters
were extracted from this large morphological database to compute
early-type fractions. There are two sources of incompleteness that
must be taken into account here. The first one is incompleteness
versus magnitude. We denote this spectroscopic completeness function
as $C_{mag}(m)$ here, and we compute it around each cluster position
by taking the ratio of the number of galaxies in the spectroscopic
SDSS catalog (database table SpecPhoto) to the number of galaxies in
the photometric SDSS catalog (database table PhotoPrimary) as a
function of Petrosian $r$ magnitude. Galaxies around a given position
on the sky were extracted from the database using the SDSS
``fGetNearbyObjEq'' function. The second source of incompleteness
comes from the spatial sampling of the SDSS fibers on the sky. Fibers
cannot be placed closer than 55\arcsec~from one another. This means
that regions with a higher surface density of targets could not be
sampled as completely as regions in the global field. The net result
for SDSS clusters is a decrease in spectroscopic sampling as a
function of decreasing clustercentric distance $R$. We can map the
spectroscopic completeness versus $R$ by computing the ratio of
galaxies in the spectroscopic and photometric SDSS catalogs as a
function of $R$. We denote this geometrical completeness function as
$C_{geom}(R)$ here. Ideally, $C_{geom}(R)$ should be computed for each
cluster because it will depend on cluster richness and apparent size
(and thus indirectly on redshift). However, in practice, there are not
enough galaxies in a single cluster to yield $C_{geom}(R)$ with
acceptable error bars. So, we opted for averaging clusters with the
same redshifts and velocity dispersions to compute $C_{geom}(R)$. We
divided the cluster list of Table~\ref{sdss-cls-list} into three
cluster groups: (1) $z < 0.06$, (2) $z > 0.06, \sigma
< 800$ km/s, and (3) $z > 0.06, \sigma > 800$ km/s. The weight
$W_{spec}(m,R)$ in the spectroscopic catalog of a galaxy with a
$r'$-band magnitude $m$ at a clustercentric $R$ is thus given by the
product ${1\over{C_{mag}(m)}} {1\over{C_{geom}(R)}}$, and the
completeness-weighted early-fraction of a SDSS cluster is then simply:

\begin{eqnarray}
f_{et}\biggl(\substack{M_{V}\leq-19.8, \\B/T\geq0.35, \\S2\leq0.075}\biggr) = {\displaystyle \sum_{\substack{i \in [M_{V}\leq-19.8,\\R\leq R_{et},\\B/T\geq0.35,\\S2\leq0.075]}} W_{spec}(m_i,R_i) \over{\displaystyle\sum_{\substack{i \in [M_{V}\leq-19.8, \\ R\leq R_{et}]}} W_{spec}(m_i,R_i)}}
\label{sdss-et}
\end{eqnarray}

In terms of spatial resolution, the ACS, SDSS and FORS2 images have
sampling of 0.68 kpc/FWHM at $z$ = 0.8 (0\farcs09 FWHM in $i$), 1.87
kpc/FWHM at $z$ = 0.07 (1\farcs4 FWHM in $g$) and 4.5 kpc/FWHM at $z$
= 0.8 (0\farcs6 FWHM in $I$) respectively. Even though the sampling of
the ACS and FORS2 images differs by a factor of seven, their limits on
$B/T$ and $S2$ for the computation of consistent early-type galaxy
fractions were quite similar. This is an indication of the robustness
of our measured structural parameters over this range of spatial
resolutions. For the sake of simplicity, we therefore adopt the ACS
limits ($(B/T)_{SDSS,g} \geq 0.35$ and $S2_{SDSS,g} \leq 0.075$) for
our SDSS early-type galaxy fractions rather than use yet another set
of limits. We can further test these limits on the catalogue of
visually classified galaxies from the SDSS North Equatorial Region of
\citet{fukugita07}. This catalogue contains Hubble T-type visual
classifications for 2253 galaxies down to a magnitude limit of $r$ =
16. If we apply our limits on $(B/T)_{SDSS,g}$ and $S2_{SDSS,g}$ to
galaxies in this catalogue, then we find that the coefficients of the
SDSS-to-visual equivalent of Equation~\ref{fet-gim2d} would be 0.88,
0.68, 0.14, and 0.014 respectively. Early-type SDSS galaxies are
therefore quantitatively selected with an ``efficiency'' comparable to
our selection from the ACS images. 

\begin{figure}
\resizebox{\hsize}{!}{\includegraphics[angle=270]{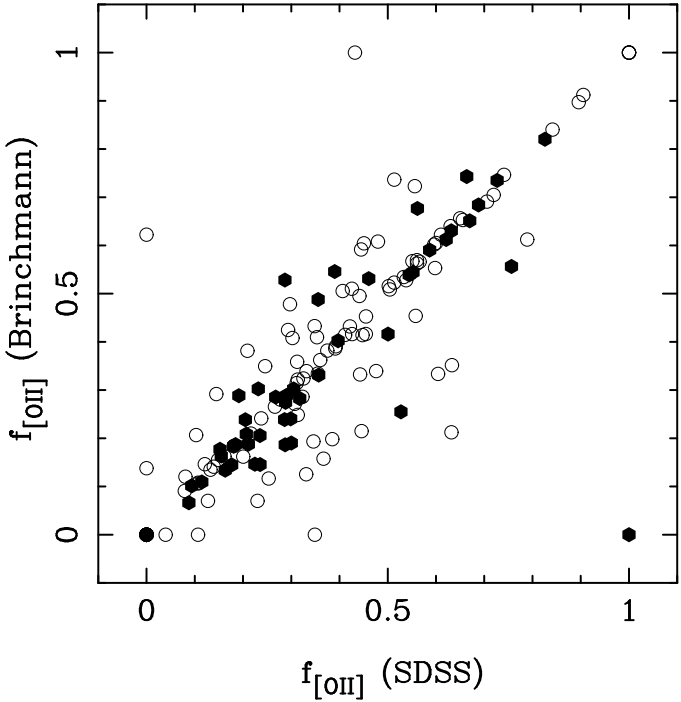}}
\caption{Comparison between fractions of [OII] emitters computed using emission-line measurements from \citet{brinchmann04} and the DR7 release. Filled and open circles are clusters with $\sigma \geq$ 600 km/s and $\sigma <$ 600 km/s respectively.}
\label{fOII-brinch_vs_sdss}
\end{figure}

The raw fractions of [OII] emitters for the 158 SDSS-C4 clusters were calculated by directly querying the SDSS database table SpecLine for the [OII]3727 and [OII]3730 equivalent widths for each confirmed cluster member, adding them together and correcting them to rest-frame by dividing by (1+$z$). The corrected [OII] fractions were then computed following exactly the same calculations (and using the same weights, the same luminosity and clustercentric radius cuts of $M_V \leq - 19.8$ and $R \leq 0.6 R_{200}$) as for the early-type fractions except that the early-type selection criteria on bulge fraction and image smoothness were simply replaced by the \citet{poggianti06} cut of EW([OII]) $\leq -3 \AA$.  In order to evaluate the importance of the errors on our equivalent widths on our determination of the fractions of [OII] emitters, we also computed [OII] fractions using equivalent widths from \citet{brinchmann04}. The two sets of equivalent widths are plotted against one another in Figure~\ref{fOII-brinch_vs_sdss}. The agreement between the two sets is excellent, and we conclude that our [OII] fractions are robust.

Table~\ref{sdss-cls-list} gives corrected early-type galaxy fractions and fractions of [OII] computed for $R \leq
0.6R_{200}$ for the 158 SDSS clusters in our local comparison sample. We included only galaxies brighter than
$M_{V,lim} = -19.8$ to avoid incompleteness in the SDSS spectroscopic
sample. This cutoff magnitude corresponds to the absolute magnitude
limits we used for our distant EDisCS clusters once passive evolution
is taken into account (see Section~\ref{efrac-HST}).

\addtocounter{table}{1}

\subsection{Theoretical Models}\label{theo-models}

Numerical simulations of dark matter haloes populated with galaxies using semi-analytical models greatly help in the interpretation of observational results. We use here the Millennium Simulation \citep[MS;][]{springel05}, and the semi-analytical code described in \citet{delucia07a}\footnote{Simulated galaxy catalogs used here are publicly available at http://www.mpa-garching.mpg.de/millennium/}. The MS followed 2160$^3$ particles of mass 8.6$\times$10$^8 h^{-1}$M$_\odot$ within a comoving box of size 500 $h^{-1}$ Mpc on a side with a spatial resolution of 5$h^{-1}$ kpc.  Early-type galaxy fractions were computed from these simulated galaxy catalogs using the following procedure. Haloes were randomly selected at three different redshifts ($z$ = 0, 0.41, 0.62) so that they were uniformly distributed in
log($M_{200}$). The final halo sample was 100 haloes at $z$ = 0, 94 haloes at $z$ = 0.41 and 92 haloes at $z$ = 0.62. For each of these haloes, all galaxies in a cubic box 6 Mpc on a side around the central galaxy were selected, and a morphological type was assigned to each model galaxy by computing the quantity $\Delta M$ = $M_{bulge} - M_{total}$ (in the rest-frame $B$-band). Galaxies with $\Delta M <$ 1.0 were considered to be "early-type". This is the same criterion as selecting real galaxies with $B/T_{FORS2} \geq 0.40$. It is important here to note that an early-type galaxy in the simulations was defined solely based on this cut in bulge fraction because the simulations do not have the resolution required to model internal fine structures such as asymmetries. Given that real, early-type galaxies were also selected according to image smoothness, one might find the early-type fractions of real clusters to be systematically lower. For each halo, the fraction of early-type galaxies within 0.6$R_{200}$ from the BCG was computed using three different projections. Furthermore, only galaxies that were within 2 Mpc from the BCG along the line of sight were included. The fractions were computed using only galaxies brighter than
$-$20.5, $-$20.1, and $-$19.8 in the rest-frame $V-$band at redshift 0.6, 0.4, and 0.0
respectively to match the limits used for the SDSS and EDisCS early-type galaxy fractions. A galaxy in the simulation was deemed to be star-forming if its star-formation rate in the last timestep of its evolution was not equal to zero.

\begin{figure*}
\resizebox{\hsize}{!}{\includegraphics[angle=90]{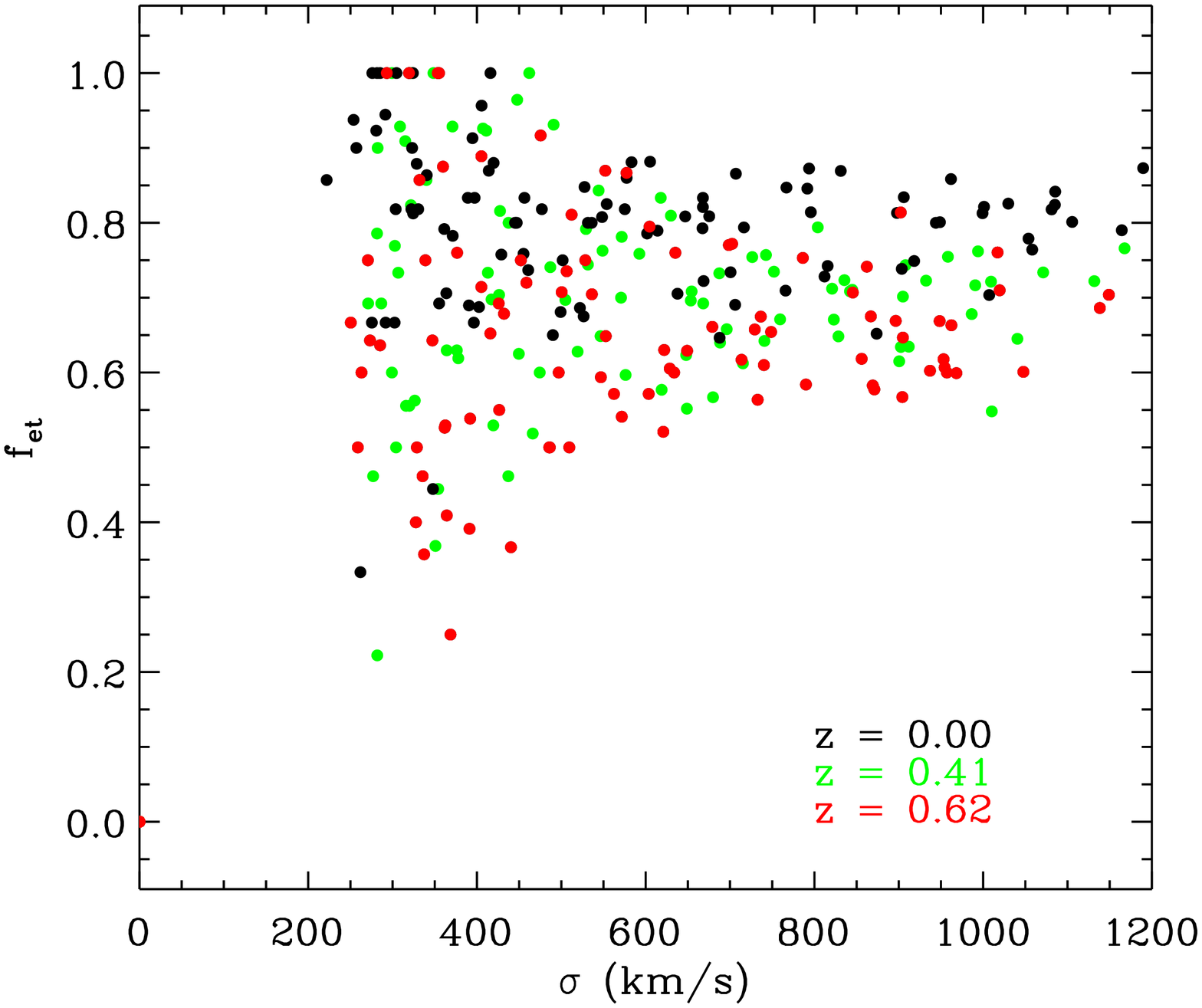},\includegraphics[angle=90]{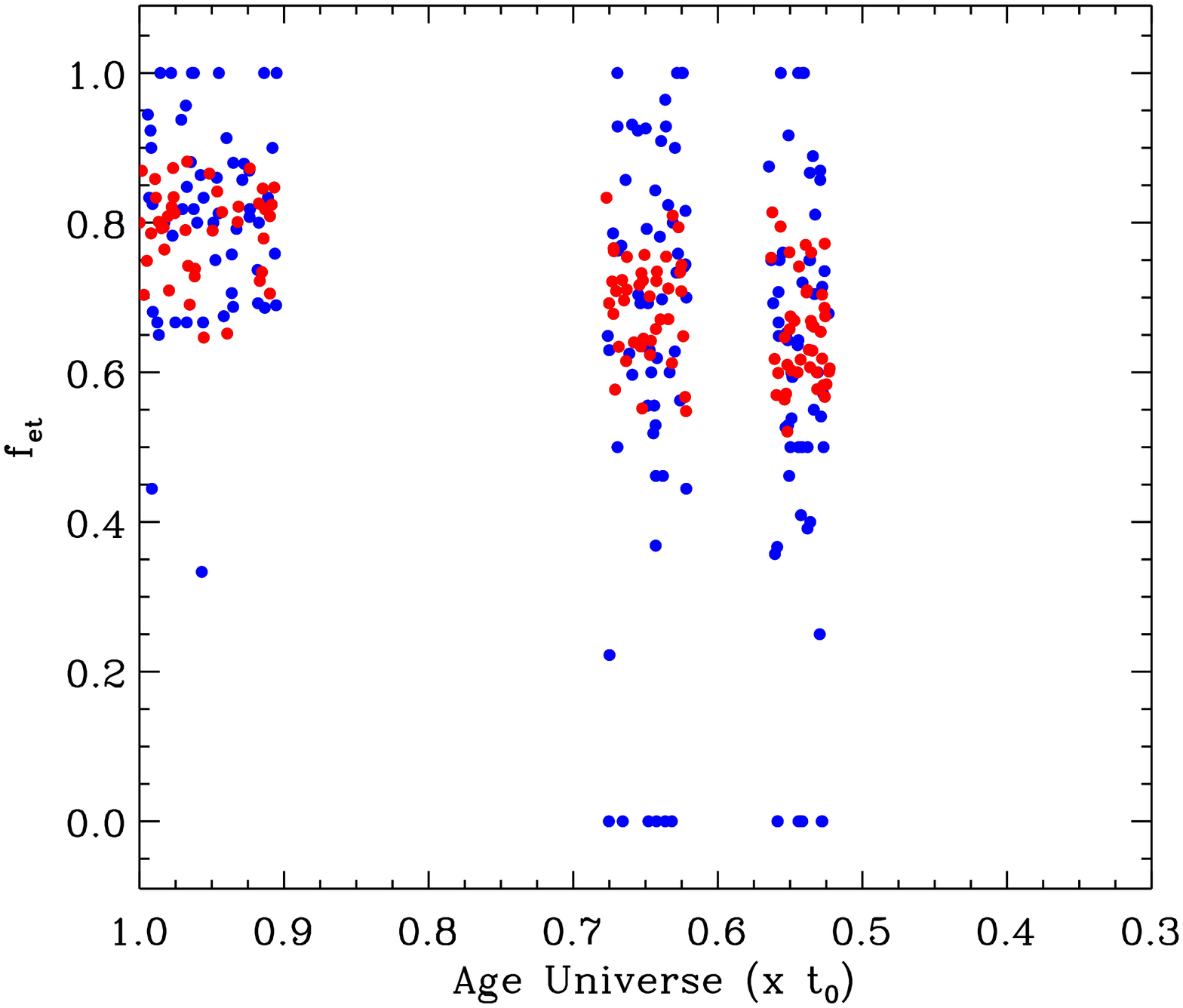}}
\resizebox{\hsize}{!}{\includegraphics[angle=90]{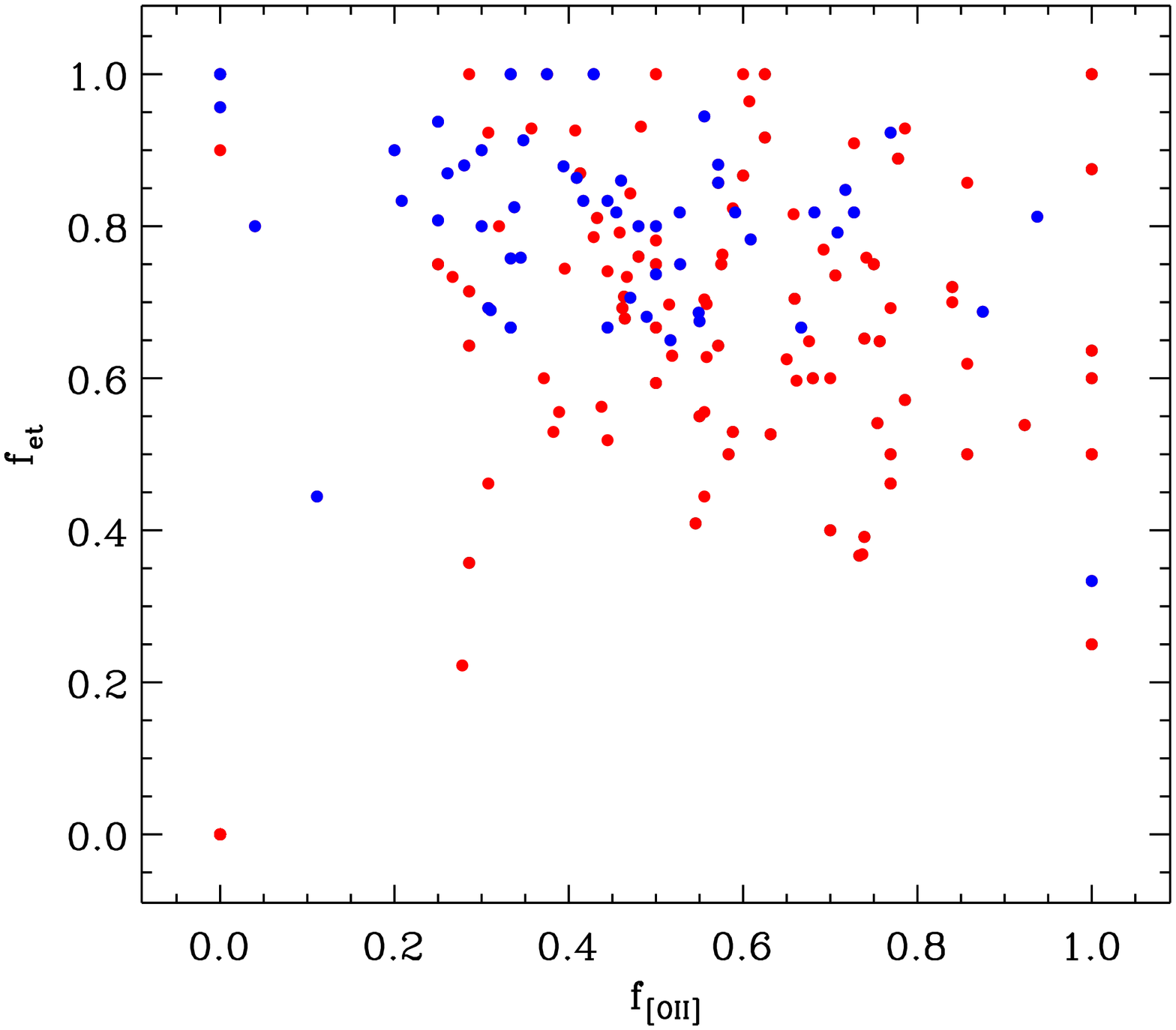},\includegraphics[angle=90]{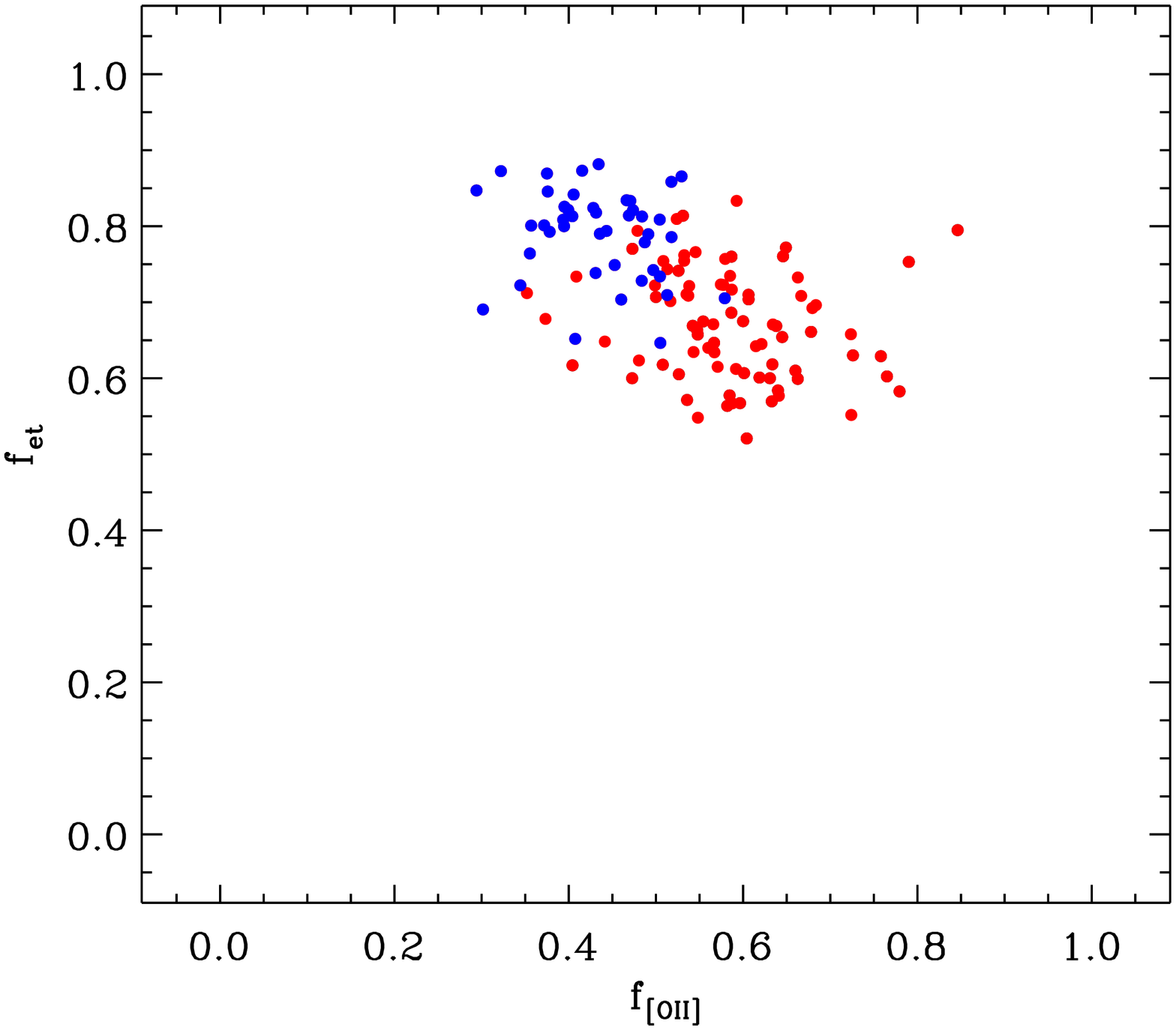}}

\caption{Early-type galaxies in Millennium Simulation dark matter haloes {\it Top, left-hand panel:}~Early-type galaxy fraction within $0.6 R_{200}$ versus cluster velocity dispersion at three different redshifts. {\it Top, right-hand panel:} Early-type galaxy fraction within $0.6 R_{200}$ versus age of the universe. Blue and red points are clusters with velocity dispersions below and above 600 km/s respectively. {\it Lower, left-hand panel:}~ Early-type galaxy fraction within $0.6 R_{200}$ versus fraction of star-forming galaxies in clusters with $\sigma <$ 600 km/s. Blue points show haloes selected at redshift zero, and all the other haloes are in red. {\it Lower, right-hand panel:}~ Early-type galaxy fraction within $0.6 R_{200}$ versus fraction of star-forming galaxies in clusters with $\sigma \geq$ 600 km/s. }
\label{theo-efraction-plot}
\end{figure*}

Figure~\ref{theo-efraction-plot} shows the resulting model early-type fractions as a function of cluster velocity dispersion, redshift and fraction of star-forming galaxies for the MS haloes. At a given redshift, there is no dependence of the early-type fraction on cluster velocity dispersion, but the scatter symmetrically increases towards both lower and higher fractions leading to a "wedge-like" distribution towards lower cluster $\sigma$'s. The early-type fractions of both low and high-mass clusters increase with decreasing redshift from $\sim$0.70 at $z = 0.65$ to $\sim$0.85 at $z$ = 0. The early-type fractions of massive clusters are anticorrelated with the fractions of star-forming galaxies: clusters at $z = 0$ have higher early-type fractions but lower fractions of star-forming galaxies. Note that the trends in Figure~\ref{theo-efraction-plot} do not agree with those shown in \citet{diaferio01} although the assumptions made about morphological transformations are very similar in the two models. In particular, the MS  shows little trend of early-type fraction with cluster velocity dispersion but a substantial trend with redshift, while Diaferio et al. found the opposite. This is likely a result of the poorer mass resolution, poorer statistics and cruder dynamical modelling of the earlier paper. 

\section{Results}\label{results}

We use here our VLT/FORS2 early-type fractions for all EDisCS clusters for the sake of uniformity.

\subsection{Early-Type Galaxy Fractions versus Cluster Velocity Dispersion and Redshift}\label{efrac-sigma}

\begin{figure*}
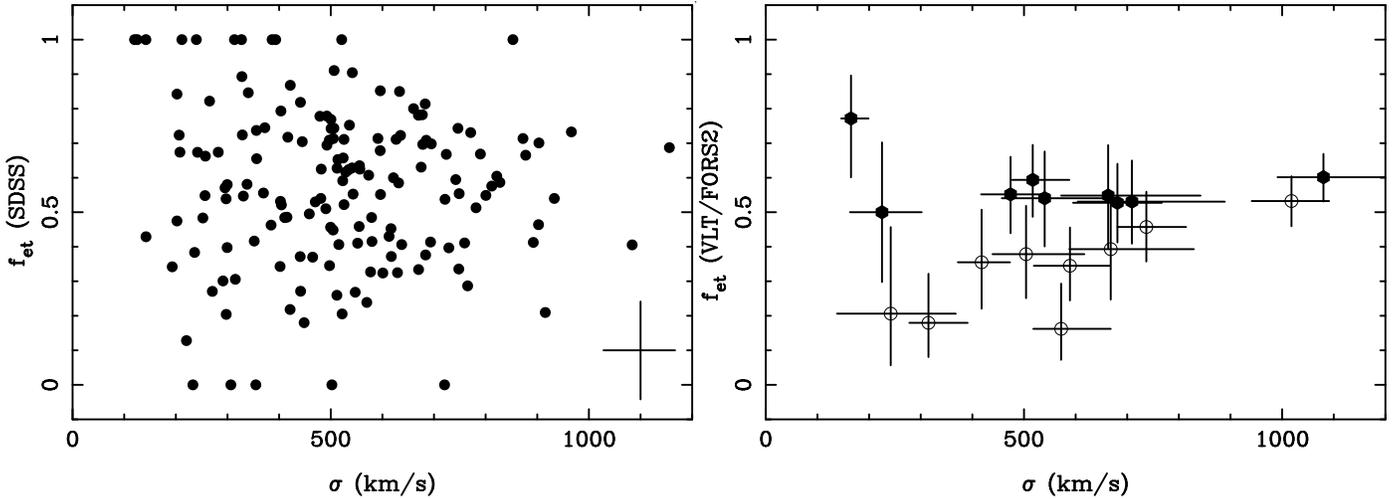

\resizebox{\hsize}{!}{\includegraphics[angle=270]{fig8a.ps},\includegraphics[angle=270]{fig8b.ps}}

\caption{Early-type galaxy fraction within $0.6 R_{200}$ versus velocity dispersion for SDSS and EDisCS clusters. Both samples have been matched in velocity dispersion. {\it Left panel:} SDSS clusters. Only typical error bars are shown in the lower right-hand corner for clarity. {\it Right panel:} Filled and open circles are mid-$z$ and high-$z$ EDisCS clusters respectively. Errors bars shown in both panels are 1$\sigma$ errors. Our VLT/FORS2 early-type fractions are used here for all EDisCS clusters for the sake of uniformity.}
\label{vlt+sdss_efrac_sigma}
\end{figure*}

Figure~\ref{vlt+sdss_efrac_sigma} shows early-type galaxy fractions
versus velocity dispersion for the SDSS and EDisCS clusters. The
early-type galaxy fractions of both cluster samples exhibit no clear trend as a function of $\sigma$. Table~\ref{spearman-tests} gives Spearman rank test results for the SDSS sample and different EDisCS subsamples. The only significant correlation between early-type fraction and velocity dispersion is found in the high-z EDisCS clusters. It only has a 2.5$\%$ chance of being due to randon sampling. Such a positive correlation was also reported in \citet{desai07} for the same cluster subsample, but it disappears when the full EDisCS sample is considered.  The lack of a significant correlation agrees well with the results for the Millennium Simulation in the top left-hand panel of Figure~\ref{theo-efraction-plot} but disagrees with the earlier theoretical results of \citet{diaferio01} which showed a trend between $f_{et}$ and $\sigma$.  A visual inspection of Figure~\ref{vlt+sdss_efrac_sigma} confirms the statistical test results. The mid-z EDisCS clusters do not show any correlation with $\sigma$ in contrast to the high-z clusters. In particular, two mid-$z$ EDisCS clusters (CL1119.3-1130 and CL1420.3-1236) with $\sigma \sim$ 200 km/s have early-type fractions similar or higher ($f_{et} \sim$ 0.5-0.8) than the most massive clusters in our sample. Interestingly, the same two clusters were found by \cite{poggianti06} to be the most outstanding outliers in the [OII] fraction - $\sigma$ relation in the sense that they have a low fraction of [OII] emitters for their mass. This is consistent with what we observe here given that early-type galaxies typically have lower [OII] emission fluxes. 

Figure~\ref{vlt+sdss_efrac_sigma} does show that there is a marked difference in the morphological content of the EDisCS and SDSS clusters.   All EDisCS $f_{et}$ values (with the exception of one cluster) are below 0.6, but half of the SDSS clusters are above this value.  The population of early-type galaxies has thus increased significantly in half of the clusters {\it of all velocity dispersions}. An increase in early-type fraction with decreasing redshift may already be visible when one compares mid-$z$ and high-$z$ EDisCS clusters. Mid-$z$ clusters around $\sigma$ = 500 km/s have $f_{et} \simeq$ 0.5 whereas the high-$z$ clusters have $f_{et} \simeq$ 0.4. This would represent a $\sim$25$\%$ increase over a time interval of 2 Gyrs. As shown in Figure~\ref{theo-efraction-plot}, the early-type fractions of clusters in the Millenium Simulation also increase with decreasing redshift in clusters of all velocity dispersions, but there is a lack of simulated clusters with $f_{et} < 0.5$ compared with the SDSS-C4 clusters. The scatter in the $f_{et}$ values of simulated clusters is also smaller than in those of real clusters. For simulated clusters at $z = 0$ with $\sigma \geq $ 600 km/s, $\sigma(f_{et})$ = 0.06 compared to $\sigma(f_{et})$ = 0.21 for SDSS clusters over the same range of velocity dispersions. Given that the mean error on the SDSS $f_{et}$ values is 0.12, the intrinsic scatter would be 0.17. This intrinsic scatter is still almost three times the scatter in the simulated clusters.

\begin{table}
\caption[]{Spearman Rank Tests Results for Early-Type Fraction versus Cluster Velocity Dispersion}
\begin{center}
\begin{tabular}{lrrr}
\hline
Cluster Sample & $N_{cl}$ & $R_s$ & $p$-value\\
(1) & (2) & (3) & (4)\\
\hline
SDSS &158 & $-$0.05 & 0.51\\
EDisCS all & 18 & 0.18 & 0.47\\
EDisCS mid-z & 9 & $-$0.11 & 0.78\\
EDisCS high-z & 9 & 0.73 & 0.025\\
\hline
\end{tabular}
\label{spearman-tests}
\end{center}
\end{table}

\begin{table}
\caption[]{Two-sample  Kolmogorov-Smirnov Test Probabilities for Early-Type Fraction versus Cluster Velocity Dispersion}
\begin{tabular}{lrrrr}
\hline
& \multicolumn{4}{c}{Cluster Sample 2}\\
 & SDSS & SDSS & EDisCS  &EDisCS  \\
Cluster Sample 1 & (All) & ($\sigma \geq$ 600) & ($\sigma <$ 600) & ($\sigma \geq$ 600) \\
\hline
SDSS ($\sigma <$ 600) &  \ldots & 0.628 & 0.190 & \ldots\\
SDSS ($\sigma \geq$ 600) & \ldots & \ldots & \ldots & 0.472\\
EDisCS  ($\sigma <$ 600) &  \ldots & \ldots & \ldots & 0.250\\
EDisCS (All) & 0.506 & \ldots & \ldots & \ldots \\
\hline
\end{tabular}
\label{ks-tests-fet}
\end{table}

\begin{figure*}
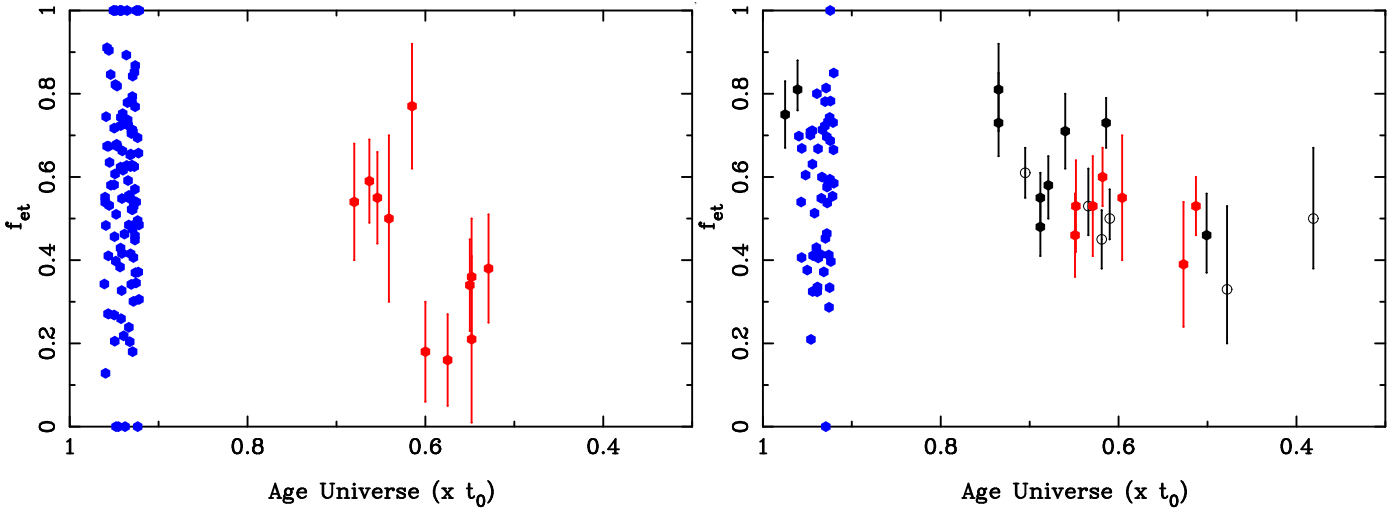

\resizebox{\hsize}{!}{\includegraphics[angle=270]{fig9a.ps},\includegraphics[angle=270]{fig9b.ps}}
\caption{Early-type galaxy fraction versus age of the universe (i.e., redshift) for clusters with $\sigma < 600$ km/s (left panel) and clusters with $\sigma \geq 600$ km/s (right panels). SDSS and EDisCS clusters are blue and red respectively, and both samples have been matched in velocity dispersion. Clusters shown in black are from the compilation of \citet{dokkum01} in which open and solid points have X-ray luminosities below and over 10$^{44.5}$ ergs s$^{-1}$ respectively. Our VLT/FORS2 early-type fractions are used here for all EDisCS clusters for the sake of uniformity.}  
\label{vlt+sdss_efrac_age}
\end{figure*}

Figure~\ref{vlt+sdss_efrac_age} shows SDSS and EDisCS early-type fractions as a function of the age of the universe (i.e., redshift). The clusters have been divided into two subgroups based on their velocity dispersions. The early-type fractions of massive ($\sigma >$ 600 km/s) EDisCS clusters (right panel) are in very good agreement with the ones in the compilation of \citet{dokkum01} which also have velocity dispersions greater than 600 km/s. The clusters at low redshift in the \citet{dokkum01} compilation suggest that there are no local clusters with low early-type fractions and hence that all clusters have uniformly increased their early-type fraction from $z \sim 0.$ to the present day. However, our SDSS cluster sample shows that this simple picture is not entirely true. While half of the SDSS clusters have higher early-type fractions than clusters at high redshift, the other half  have early-type fractions equal or even lower than the EDisCS clusters. The same holds true for the low mass clusters (left-hand panel). The scatter in $f_{et}$ ($<$ 0.1) in high-mass EDisCS clusters does appear to be considerably less that the scatter seen in low-mass clusters.

\begin{figure*}
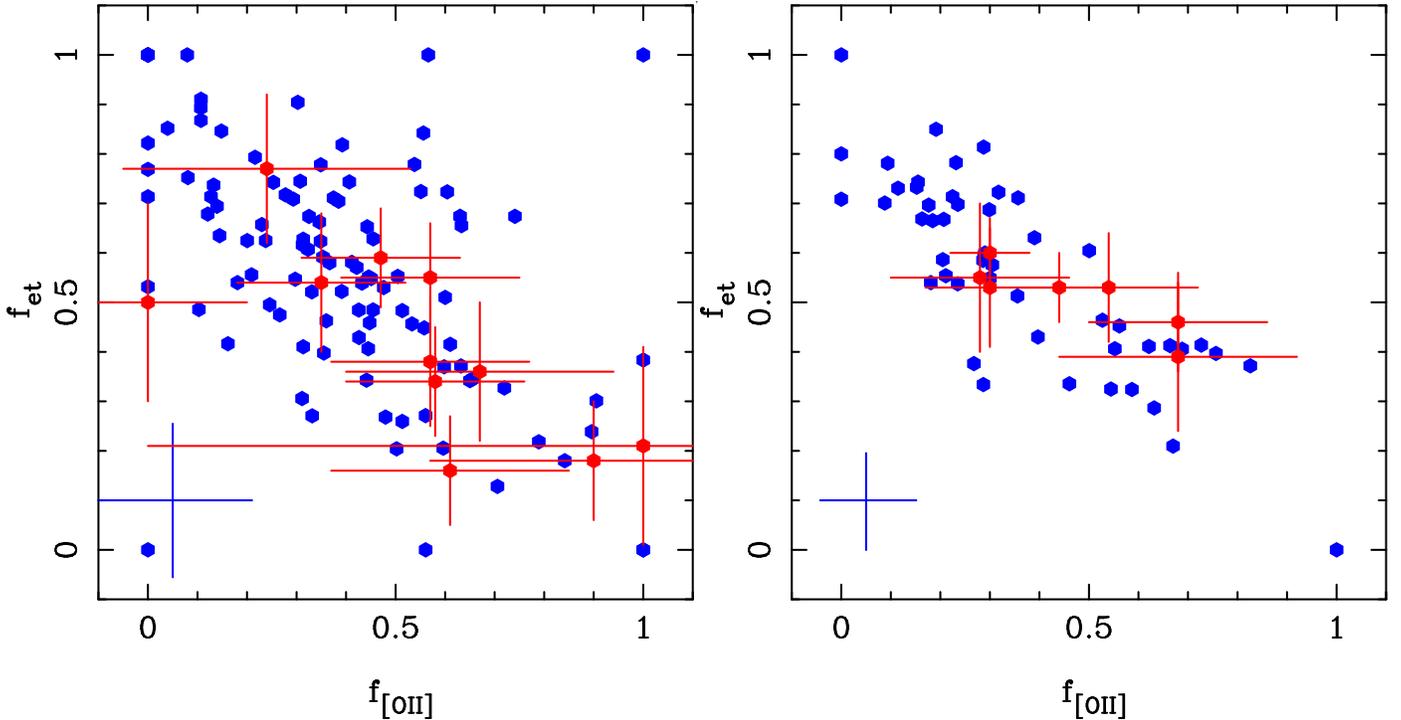

\resizebox{\hsize}{!}{\includegraphics[angle=270]{fig10a.ps},\includegraphics[angle=270]{fig10b.ps}}
\caption{Early-type galaxy fraction versus [OII] emitter fraction for clusters with $\sigma < 600$ km/s (left panel) and clusters with $\sigma \geq 600$ km/s (right panel). SDSS and EDisCS clusters are shown in blue and red respectively, and both samples have been matched in velocity dispersion. Only typical error bars are shown for the SDSS clusters in the lower right-hand corner for clarity. Our VLT/FORS2 early-type fractions are used here for all EDisCS clusters for the sake of uniformity.}
\label{vlt+sdss_fet_fOII}
\end{figure*}

The lack of a clear trend in early-type fraction with redshift in the right-hand panel of Figure~\ref{vlt+sdss_efrac_age} is in disagreement with the Millennium Simulation prediction in the top right-hand panel of Figure~\ref{theo-efraction-plot}. There is a clear deficit of clusters with low early-type fraction  at low redshift in the Millenium Simulation compared with our SDSS sample.

\subsection{Early-Type Galaxy Fractions versus Fractions of [OII] Emitters}\label{efrac-OII}
The link between star formation and morphological transformation and
its evolution as a function of redshift provides more clues on the
processes driving galaxy morphology in local and distant
clusters. The fractions of galaxies
with [OII] emission in the EDisCS clusters were computed as in \citet{poggianti06}~using the same
absolute magnitude limits and the same prescriptions for correcting magnitude and geometric incompletness, but the clustercentric radius cut was changed to match the one used for the early-type
fractions in this paper ($R_{et} \leq 0.6R_{200}$). The two datasets are
therefore directly comparable. Figure~\ref{vlt+sdss_fet_fOII} shows
$f_{et}$ versus $f_{[OII]}$ with our local and distant samples again
divided according to velocity dispersion. Table~\ref{spearman-tests-fo2} gives Spearman test results between $f_{et}$ and  $f_{[OII]}$. There is a strong correlation between  $f_{et}$ versus $f_{[OII]}$ in both SDSS and EDisCS cluster samples irrespective of cluster velocity dispersion. The EDisCS clusters lie within the envelopes defined by the SDSS clusters. There is no offset between the zeropoints of the correlations at low and high redshift. However, as demonstrated by
\citet{poggianti06}, the star formation activity (parametrized by
$f_{[OII]}$) has decreased in all environments from $z \sim 0.75$ to
$z \sim 0.08$. This is confirmed by the K-S test results in Table~\ref{ks-tests-fo2}. The probabilities that the EDisCS and SDSS clusters are drawn from the same parent $f_{[OII]}$ distribution are only 0.026, 0.005 and 0.046 for the whole samples, low $\sigma$ and high $\sigma$ subsamples respectively.

The $f_{et}$ versus $f_{[OII]}$ values for clusters from the Millenium Simulation (Figure~\ref{theo-efraction-plot}) are quite different from the observations. Low $\sigma$ MS clusters at low and high redshifts are confined to high $f_{et}$ and $f_{OII}$ values with no apparent correlation. There is only a handful of clusters with low values for both $f_{et}$ and $f_{OII}$. The high $\sigma$ MS clusters are found in a very limited range of $f_{et}$ and $f_{OII}$ values ($0.35 < f_{OII} < 0.75$, $0.6 < f_{et} < 0.85$).

\begin{table}
\caption[]{Spearman Rank Tests Results for Early-Type Fraction versus Fraction of [OII] Emitters}
\begin{center}
\begin{tabular}{lccc}
\hline
Cluster Sample & $N_{cl}$ & $R_s$ & $p$-value\\
(1) & (2) & (3) & (4)\\
\hline
SDSS (All) & 158 & $-$0.63 & 4.33$\times 10^{-19}$\\
SDSS ($\sigma < 600)$ & 108 & $-$0.57 & 1.50$\times 10^{-10}$\\
SDSS ($\sigma \geq 600)$ & 50 & $-$0.77 & 6.04$\times 10^{-11}$\\
EDisCS (All) & 18 & $-$0.74 & 0.00043\\
EDisCS ($\sigma < 600)$  & 11 & $-$0.78 & 0.0043\\
EDisCS ($\sigma \geq 600)$  & 7 & $-$0.77 & 0.0438\\
\hline
\end{tabular}
\label{spearman-tests-fo2}
\end{center}
\end{table}

\begin{table}
\caption[]{Two-sample  Kolmogorov-Smirnov Test Probabilities for [OII] Emitter Fraction versus Cluster Velocity Dispersion}
\begin{tabular}{lrrrr}
\hline
& \multicolumn{4}{c}{Cluster Sample 2}\\
 & SDSS & SDSS & EDisCS  &EDisCS  \\
Cluster Sample 1 & (All) & ($\sigma \geq$ 600) & ($\sigma <$ 600) & ($\sigma \geq$ 600) \\
\hline
SDSS ($\sigma <$ 600) &  \ldots & 0.090 & 0.005 & \ldots\\
SDSS ($\sigma \geq$ 600) & \ldots & \ldots & \ldots & 0.046\\
EDisCS  ($\sigma <$ 600) &  \ldots & \ldots & \ldots & 0.761\\
EDisCS (All) & 0.026 & \ldots & \ldots & \ldots \\
\hline
\end{tabular}
\label{ks-tests-fo2}
\end{table}

\section{Discussion}\label{discussion}

In order to fully understand possible evolutionary trends observed
here, it is important to determine how cluster velocity dispersion
changes with redshift as a result of the hierarchical growth of structures. Are we looking at similar clusters when we
focus on the same range of velocity dispersions in the SDSS and EDisCS
clusters? \citet{poggianti06} looked at the mean change in $\sigma$
between $z$ = 0 and $z$ = 0.76 using a sample of 90 haloes from the
Millennium Simulation uniformly distributed in
log(mass) between 5 $\times$ 10$^{12}$ and 5 $\times$ 10$^{15}$
M$_{\odot}$. Their Figure 8 shows how $\sigma$ evolves over that
redshift interval. For example, a $z$ = 0 cluster with $\sigma$ = 900
km/s would typically have $\sigma \sim$ 750 km/s at $z$ = 0.76. This evolution
is not sufficient to introduce biases in our analysis here. Indeed,
selecting clusters with $\sigma \geq$ 600 km/s, say, at either $z$ = 0
or $z$ = 0.76 would keep nearly all the same clusters. Measured velocity dispersions may exhibit a large scatter with respect to the true halo mass particularly for low-mass clusters. The velocity dispersions for the SDSS and EDisCS clusters were calculated in a very similar way in order to minimize any biases. Velocity dispersions calculated from a small number of cluster members may be overestimates of the true cluster mass. Table~\ref{clsample} lists 1103.7-1245b as the cluster with the lowest number of members ($N$ = 11). In order to check the robustness of our results, we re-ran our analyses by excluding SDSS clusters in Table~\ref{sdss-cls-list} with $N < 10$ for which velocity dispersions may be less reliable and found that our results remained unchanged.

\citet{poggianti06} proposed a scenario in which two channels are responsible for the production of passive galaxies in clusters, and others \citep{faber07,brown07} have proposed a similar scenario for the migration of galaxies from the "blue cloud" to the red sequence. "Primordial passive galaxies" are composed of galaxies whose stars all formed at very high redshift ($z >$ 2) over a short timescale. These galaxies have been observed in clusters up and beyond $z = 1$, and they largely comprise luminous ellipticals. "Quenched passive galaxies" have had a more extended period of star formation activity, and their star formation has been quenched after their infall into dense cluster environments. These quenched passive galaxies would then suffer the effects of cluster processes such as ram pressure stripping, harassment, strangulation and mergers to become S0 and earlier type galaxies. A key point of this scenario is that processes affecting morphology and star formation activity operate on different timescales as shown recently for the EDisCS sample by \citet{sanchez09}. There is good evidence that star formation is quenched in galaxies over timescales of 1-3 Gyr after they have entered the cluster environment \citep{poggianti99,poggianti06} whereas morphological transformation through mergers and harassment can take longer \citep[$\sim$ 5 Gyr,][]{moore98}. The best example of this is the fact that the vast majority of post-starburst galaxies in distant clusters, those that have had their star formation activity terminated during the last Gyr, still retain a spiral morphology \citep{poggianti99}. Such a two-channel scenario would naturally explain observations indicating that the elliptical galaxy fraction actually remains constant with redshift while the S0 fraction rises with decreasing redshift \citep{dressler97,fasano00,desai07}. Unfortunately, the VLT/FORS2 images do not have sufficient spatial resolution to disentangle E and S0 galaxies as mentioned in Section~\ref{efrac-defn} to determine the exact contribution from  each channel. We can therefore only study the overall production of early-type galaxies, but it should exhibit different behaviors with cluster global properties depending on the process(es) dominating it. Given our quantitative definition of an early-type galaxy based on bulge fraction and image smoothness, there are essentially two ways to transform late-type galaxies into early-type ones: 1) processes such as collisions and harassment that can fundamentally alter the structure of a galaxy by forming bulges and/or destroying disks and 2) quenching processes that can extinguish star forming regions responsible for some of the galaxy image asymmetries and also cause a fading of the disks.

Applying the \citet{poggianti06} scenario to our results,  the "threshold" in $f_{et}$ values in our high redshift clusters (Figures~\ref{vlt+sdss_efrac_sigma} and~\ref{vlt+sdss_efrac_age}) could be explained by a population of primordial passive galaxies that formed at even higher redshifts. Most of our high redshift clusters have early-type fractions in the range 0.3-0.6 with no correlation with cluster velocity dispersion. Are these early-type fractions indeed consistent with a populations of primordial passive galaxies? Calculations done in \citet{poggianti06} show that the fraction of galaxies at $z = 0.6$ that were present in haloes with masses greater than 3$\times$10$^{12}$ M$_\odot$ at $z$ = 2.5 is 0.4$\pm$0.2. These primordial passive galaxies can therefore account for at least 2/3 (if not all) of the early-type populations in high redshift clusters, and their high formation redshift would explain the lack of dependence of $f_{et}$ on cluster velocity dispersion.

One of our main results is that the early-type fractions of galaxy clusters increase from $z = 0.6 - 0.8$ to $z\sim 0.08$ in clusters of all velocity dispersions. What kind of morphological transformation process(es) can lead to such an evolution? Collisions and harassment both depend on galaxy-galaxy interactions and the time a galaxy has spent within the cluster environment. Cluster velocity dispersion influences the number of interactions and their duration. Higher velocity dispersions in more massive clusters yield more interactions per unit time $N$ but with shorter durations $\Delta t$ in a given time interval. One might therefore expect to see a peak in early-type type fraction at the cluster velocity dispersion where the product N$\Delta t$ is maximized. No such peak is seen in our clusters. Ram-pressure stripping is expected to go as ($n_{ICM} v_{gal}^{2.4})/\dot{M}_{rep}$ \citep{gaetz87} with $n_{ICM}$, $v_{gal}$ and $\dot{M}_{rep}$ being the density of the ICM, the velocity of the galaxies within the ICM and the rate at which galaxies can replenish their gas respectively. The fraction of passive galaxies should therefore be a relatively strong function of cluster velocity dispersion if quenching by ram pressure stripping is the dominant process. The number of post-starburst galaxies in EDisCS clusters does correlate with cluster velocity dispersion \citep{poggianti09a}, but the uniform increase in early-type fractions at all cluster velocity dispersions observed going from EDisCS to SDSS clusters is not consistent with the intracluster medium being the main cause of the changes in cluster morphological content.

Even though the early-type and [OII] emitter fractions in EDisCS and SDSS clusters show no correlation with cluster velocity dispersion \citep[][and this work]{poggianti06}, there is a very strong correlation between $f_{et}$ and $f_{OII}$. This correlation is seen at both low and high cluster masses as well as at both low and high redshifts. Morphology and star formation therefore appear to be closely linked with one another over a wide range of environments and times. However, different structural transformation and quenching processes are thought to operate over different timescales \citep[e.g.,][]{sanchez09}. Timescales range from 1-2 Gyr (based on typical cluster crossing times) for truncating star formation to 3-5 Gyr for totally extinguishing star formation in newly accreted galaxies \citep{poggianti06,tonnesen09}. Looking at the evolution of EDisCS cluster red-sequence galaxies over 2 Gyr (from $z = 0.75$ to $z$ = 0.45), \citet{sanchez09} found that morphological transformation and quenching of star formation indeed appeared to not be simultaneous. As noted in Section~\ref{efrac-sigma}, the early-type fractions of mid-$z$ EDisCS clusters may be $\sim$25$\%$ higher than the ones of high-$z$ clusters. This change would therefore have taken place over a 2 Gyr interval in our adopted cosmology.  However, the time baseline here between SDSS and EDisCS clusters is almost 6 Gyr, and, unfortunately, this is ample time to erase any difference arising from different timescales in the link between morphology and star formation.

The lack of dependence of morphology and star formation on global cluster properties such as velocity dispersion raises the question of whether changes in galaxy properties are driven by more local effects or whether they occur outside of the cluster environment. Recent work \citep{poggianti08,park09,bamford09,ellison09} have re-emphasized the strong link between galaxy properties and local galaxy density rather than cluster membership. Galaxy properties are seen to change at densities around 15-40 galaxies Mpc$^{-2}$ or projected separations of 20-30$h^{-1}$ kpc. Others \citep[e.g.,][]{kautsch08,wilman09} have suggested that the galaxy group environment might be more conducive to galaxy transformation. Our observed evolution in early-type fraction as a function of redshift and the strong correlation between morphology and star formation at all cluster masses would support the idea that cluster membership is of lesser importance than other variables such as local density in determining galaxy properties.

The properties of simulated clusters from the Millenium Simulation compare well with those of EDisCS and SDSS clusters. Their early-type fractions also show no dependence with cluster velocity dispersion in contrast to previous theoretical work \citep[e.g.][]{diaferio01} but in agreement with observations. However, there is a definite lack of MS clusters with low early-type fractions at $z$ = 0 compared to the SDSS sample. It is important here to note that an early-type galaxy in the simulations was defined solely based on its bulge fraction because the simulations do not have the resolution required to model internal fine structures such as asymmetries. Given that real, early-type galaxies were also selected according to image smoothness, one would expect the early-type fractions of real clusters to be systematically lower. However, half of the SDSS clusters have low early-type fractions not seen in the simulations at $z$ = 0, and such a large discrepancy could only be explained by a significant population of real bulge-dominated galaxies with relatively large asymmetries. It is more likely that bulge formation in the simulations may be too efficient. The scatter in $f_{et}$ values for the simulated clusters with $\sigma \geq $ 600 km/s is also nearly three times smaller than observed in the real clusters (Section~\ref{efrac-sigma}) which may indicate that the models may not include the right mixture of evolutionary processes at work on real galaxies. High-mass simulated clusters show a correlation between early-type fraction and star-forming fraction (albeit over narrower ranges than observed), but the correlation is not seen in the low-mass simulated clusters. This may be understood by high mass clusters having been formed long enough for evolutionary processes to have had enough time to act on galaxies to modify their properties whereas this is not necessarily the case for low-mass clusters. The fact that the correlation is observed in both low- and high-mass real clusters may be an indication that processes giving rise to the correlation may be more efficient (or altogether different) than modelled. It is also important to keep in mind here that the properties of a galaxy in these models are essentially driven by the mass of its parent halo.

\section{Summary}\label{conclusions}
We have presented quantitative morphologies measured from PSF-convolved, 2D bulge+disk decompositions of cluster and field galaxies on deep VLT/FORS2 images of eighteen, optically-selected galaxy clusters at $0.45 < z < 0.80$ observed as part of the ESO Distant Cluster Survey. The morphological content of these clusters was characterized by the early-type fraction within a clustercentric radius of 0.6$R_{200}$, and early-type galaxies were selected based on bulge fraction and image smoothness. We showed a very good agreement between quantitative and visual galaxy classifications. We used a set of 158 clusters extracted from the Sloan Digital Sky Survey matched in velocity dispersion to our EDisCS sample and analyzed exactly in the same way to provide a robust comparison baseline and to control systematics.  We studied trends in early-type fraction as a function of cluster mass and redshift. We also explored the link between morphology and star formation by comparing early-type fractions to the fractions of [OII] emitters in our clusters. Our main results are:

1. The early-type fractions of the SDSS and EDisCS clusters exhibit no clear trend as a function of cluster velocity dispersion.

2. Mid-$z$ EDisCS clusters around $\sigma$ = 500 km/s have $f_{et} \simeq$ 0.5 whereas high-$z$ EDisCS clusters have $f_{et} \simeq$ 0.4. This represents a $\sim$25$\%$ increase over a time interval of 2 Gyrs.

3. There is a marked difference in the morphological content of the EDisCS and SDSS samples. None of the EDisCS clusters have an early-type fraction greater than 0.6 whereas half of the SDSS clusters lie above this value. {\it This difference is seen in clusters of all velocity dispersions (i.e., masses)}.

4. There is a strong and clear correlation between morphology and star formation activity in the sense that decreasing fractions of [OII] emitters are tracked by increasing early-type fractions. This correlation holds in both low and high cluster masses as well as at both low and high redshift.

5. The early-type fractions of clusters drawn from the Millennium Simulation \citep{springel05} using the galaxy formation model of \citet{delucia07a} also show no clear dependence on cluster velocity dispersion. However, at $z$ = 0, they are not enough simulated clusters with low early-type fractions compared to the SDSS cluster sample.  While high-mass simulated clusters show a correlation between early-type fraction and star-forming fraction (albeit over narrower ranges than observed), this correlation is not seen in the low-mass simulated clusters in contrast to the real ones.

Our results pose an interesting challenge to structural transformation and star formation quenching processes that strongly depend on the global cluster environment (e.g., a dense ICM) and suggest that cluster membership may be of lesser importance than other variables in determining galaxy properties. 

\begin{acknowledgements}
We are thankful to the anonymous referee for suggestions that greatly contributed this paper. We have benefitted from the generosity of the ESO/OPC.  G. R. thanks Special Research Area No 375 of the German
Research Foundation for financial support. The Millennium Simulation databases used in this paper and the web applications providing access to them were constructed as part of the activities of the German Astrophysical Virtual Observatory. Funding for the creation and distribution of the
SDSS Archive has been provided by the Alfred P. Sloan Foundation, the
Participating Institutions, the National Aeronautics and Space
Administration, the National Science Foundation, the U.S. Department
of Energy, the Japanese Monbukagakusho, and the Max Planck
Society. The SDSS Web site is http://www.sdss.org/.The SDSS is managed
by the Astrophysical Research Consortium (ARC) for the Participating
Institutions. The Participating Institutions are The University of
Chicago, Fermilab, the Institute for Advanced Study, the Japan
Participation Group, The Johns Hopkins University, the Korean
Scientist Group, Los Alamos National Laboratory, the
Max-Planck-Institute for Astronomy (MPIA), the Max-Planck-Institute
for Astrophysics (MPA), New Mexico State University, University of
Pittsburgh, University of Portsmouth, Princeton University, the United
States Naval Observatory, and the University of Washington. The Dark Cosmology Centre is funded by the Danish National Research Foundation.
\end{acknowledgements}

\longtab{4}{
\begin{longtable}{cccccccccc}
\caption{\label{sdss-cls-list}Velocity-dispersion-matched sample of 158 SDSS clusters in order of decreasing velocity dispersion}\\
\hline\hline
SDSS C4\footnotemark & z & $N$\footnotemark & $\sigma_v$ & $R_{200}$ & $M_{cl}$ &  \multicolumn{4}{c}{$R \leq 0.6 R_{200}$} \\ 
ID & & & (km/s) & (Mpc) & ($10^{15}$M$_\odot$) & $f_{et,raw}$ & $f_{et,corr}$ & $f_{[OII],raw}$ & $f_{[OII],corr}$ \\ 
(1) & (2) & (3) & (4) & (5) & (6) & (7) & (8) & (9) & (10) \\ 
\hline
\endfirsthead
\caption{continued.}\\
\hline\hline
SDSS C4 & z & $N$ & $\sigma_v$ & $R_{200}$ & $M_{cl}$ &  \multicolumn{4}{c}{$R \leq 0.6 R_{200}$} \\ 
ID & & & (km/s) & (Mpc) & ($10^{15}$M$_\odot$) & $f_{et,raw}$ & $f_{et,corr}$ & $f_{[OII],raw}$ & $f_{[OII],corr}$ \\ 
(1) & (2) & (3) & (4) & (5) & (6) & (7) & (8) & (9) & (10) \\ 
\hline
\endhead
\hline
\endfoot
3004 & 0.0801 & 199 & 1156$\pm$61 & 2.75 & 2.550 & 0.69$\pm$0.04 & 0.69$\pm$0.04 & 0.27$\pm$0.04 & 0.30$\pm$0.04 \\ 
2035 & 0.0652 & 77 & 1084$\pm$114 & 2.60 & 2.117 & 0.38$\pm$0.11 & 0.41$\pm$0.12 & 0.67$\pm$0.11 & 0.69$\pm$0.11 \\ 
1004 & 0.0774 & 127 & 966$\pm$59 & 2.30 & 1.491 & 0.77$\pm$0.04 & 0.73$\pm$0.05 & 0.14$\pm$0.04 & 0.15$\pm$0.04 \\ 
2026 & 0.0444 & 147 & 933$\pm$71 & 2.26 & 1.364 & 0.56$\pm$0.10 & 0.54$\pm$0.10 & 0.18$\pm$0.07 & 0.18$\pm$0.07 \\ 
2159 & 0.0563 & 44 & 915$\pm$69 & 2.20 & 1.281 & 0.28$\pm$0.12 & 0.21$\pm$0.12 & 0.65$\pm$0.13 & 0.67$\pm$0.13 \\ 
2013 & 0.0556 & 160 & 903$\pm$56 & 2.18 & 1.231 & 0.70$\pm$0.06 & 0.70$\pm$0.06 & 0.08$\pm$0.03 & 0.09$\pm$0.04 \\ 
3347 & 0.0759 & 30 & 902$\pm$102 & 2.15 & 1.216 & 0.50$\pm$0.12 & 0.46$\pm$0.12 & 0.50$\pm$0.12 & 0.53$\pm$0.12 \\ 
3500 & 0.0783 & 18 & 892$\pm$113 & 2.13 & 1.174 & 0.50$\pm$0.14 & 0.41$\pm$0.14 & 0.59$\pm$0.14 & 0.66$\pm$0.14 \\ 
1126 & 0.0843 & 57 & 878$\pm$77 & 2.08 & 1.113 & 0.62$\pm$0.08 & 0.67$\pm$0.08 & 0.17$\pm$0.06 & 0.18$\pm$0.07 \\ 
3028 & 0.0704 & 121 & 872$\pm$54 & 2.09 & 1.101 & 0.69$\pm$0.08 & 0.71$\pm$0.08 & 0.21$\pm$0.07 & 0.22$\pm$0.08 \\ 
1389 & 0.0801 & 16 & 853$\pm$134 & 2.03 & 1.024 & 0.71$\pm$0.26 & 1.00$\pm$0.26 & 0.29$\pm$0.26 & 0.00$\pm$0.26 \\ 
1048 & 0.0774 & 75 & 828$\pm$78 & 1.97 & 0.937 & 0.60$\pm$0.07 & 0.59$\pm$0.07 & 0.22$\pm$0.06 & 0.20$\pm$0.06 \\ 
3016 & 0.0497 & 101 & 822$\pm$54 & 1.98 & 0.929 & 0.60$\pm$0.08 & 0.60$\pm$0.08 & 0.47$\pm$0.08 & 0.50$\pm$0.08 \\ 
2002 & 0.0762 & 102 & 812$\pm$49 & 1.94 & 0.885 & 0.63$\pm$0.06 & 0.58$\pm$0.06 & 0.28$\pm$0.05 & 0.30$\pm$0.06 \\ 
1002 & 0.0690 & 90 & 800$\pm$56 & 1.92 & 0.851 & 0.57$\pm$0.07 & 0.55$\pm$0.07 & 0.28$\pm$0.06 & 0.30$\pm$0.06 \\ 
1025 & 0.0451 & 66 & 790$\pm$54 & 1.91 & 0.827 & 0.68$\pm$0.09 & 0.67$\pm$0.10 & 0.15$\pm$0.07 & 0.16$\pm$0.08 \\ 
3084 & 0.0607 & 65 & 781$\pm$70 & 1.88 & 0.795 & 0.52$\pm$0.10 & 0.51$\pm$0.10 & 0.37$\pm$0.09 & 0.36$\pm$0.09 \\ 
1044 & 0.0837 & 60 & 771$\pm$79 & 1.83 & 0.755 & 0.72$\pm$0.07 & 0.73$\pm$0.07 & 0.12$\pm$0.05 & 0.11$\pm$0.05 \\ 
2074 & 0.0787 & 21 & 765$\pm$99 & 1.82 & 0.740 & 0.24$\pm$0.11 & 0.29$\pm$0.11 & 0.63$\pm$0.12 & 0.63$\pm$0.12 \\ 
2050 & 0.0588 & 62 & 759$\pm$59 & 1.83 & 0.730 & 0.42$\pm$0.19 & 0.41$\pm$0.36 & 0.58$\pm$0.19 & 0.62$\pm$0.33 \\ 
1058 & 0.0831 & 68 & 749$\pm$63 & 1.78 & 0.692 & 0.58$\pm$0.07 & 0.55$\pm$0.07 & 0.19$\pm$0.05 & 0.21$\pm$0.06 \\ 
1401 & 0.0643 & 28 & 748$\pm$97 & 1.79 & 0.696 & 0.40$\pm$0.13 & 0.34$\pm$0.12 & 0.47$\pm$0.13 & 0.46$\pm$0.13 \\ 
1001 & 0.0794 & 82 & 746$\pm$58 & 1.78 & 0.687 & 0.74$\pm$0.06 & 0.74$\pm$0.06 & 0.15$\pm$0.05 & 0.15$\pm$0.05 \\ 
2015 & 0.0797 & 59 & 742$\pm$77 & 1.77 & 0.675 & 0.61$\pm$0.08 & 0.59$\pm$0.08 & 0.29$\pm$0.07 & 0.29$\pm$0.07 \\ 
1276 & 0.0810 & 17 & 729$\pm$79 & 1.73 & 0.639 & 0.43$\pm$0.13 & 0.40$\pm$0.13 & 0.72$\pm$0.12 & 0.76$\pm$0.12 \\ 
3065 & 0.0649 & 83 & 724$\pm$54 & 1.74 & 0.631 & 0.66$\pm$0.08 & 0.67$\pm$0.08 & 0.23$\pm$0.07 & 0.21$\pm$0.07 \\ 
1069 & 0.0764 & 59 & 721$\pm$69 & 1.72 & 0.621 & 0.53$\pm$0.09 & 0.54$\pm$0.09 & 0.26$\pm$0.08 & 0.24$\pm$0.08 \\ 
2069 & 0.0746 & 22 & 720$\pm$119 & 1.72 & 0.619 & 0.50$\pm$0.34 & 0.00$\pm$0.34 & 0.50$\pm$0.34 & 1.00$\pm$0.34 \\ 
2001 & 0.0417 & 125 & 695$\pm$52 & 1.68 & 0.564 & 0.68$\pm$0.09 & 0.70$\pm$0.09 & 0.24$\pm$0.08 & 0.24$\pm$0.08 \\ 
3630 & 0.0682 & 18 & 693$\pm$78 & 1.66 & 0.553 & 0.44$\pm$0.16 & 0.41$\pm$0.17 & 0.68$\pm$0.15 & 0.73$\pm$0.16 \\ 
1291 & 0.0557 & 25 & 685$\pm$84 & 1.65 & 0.537 & 0.50$\pm$0.34 & 0.71$\pm$0.34 & 0.50$\pm$0.34 & 0.00$\pm$0.34 \\ 
3018 & 0.0517 & 85 & 684$\pm$52 & 1.65 & 0.535 & 0.36$\pm$0.08 & 0.38$\pm$0.09 & 0.27$\pm$0.08 & 0.27$\pm$0.08 \\ 
3404 & 0.0750 & 19 & 683$\pm$101 & 1.63 & 0.527 & 0.78$\pm$0.11 & 0.81$\pm$0.12 & 0.30$\pm$0.13 & 0.29$\pm$0.13 \\ 
1041 & 0.0758 & 62 & 678$\pm$68 & 1.62 & 0.516 & 0.70$\pm$0.08 & 0.70$\pm$0.08 & 0.19$\pm$0.06 & 0.18$\pm$0.07 \\ 
1372 & 0.0804 & 34 & 677$\pm$51 & 1.61 & 0.513 & 0.74$\pm$0.11 & 0.78$\pm$0.11 & 0.26$\pm$0.11 & 0.23$\pm$0.11 \\ 
3055 & 0.0581 & 29 & 675$\pm$82 & 1.62 & 0.513 & 0.64$\pm$0.10 & 0.63$\pm$0.10 & 0.36$\pm$0.10 & 0.39$\pm$0.10 \\ 
3027 & 0.0737 & 38 & 670$\pm$74 & 1.60 & 0.498 & 0.76$\pm$0.09 & 0.78$\pm$0.09 & 0.11$\pm$0.06 & 0.09$\pm$0.07 \\ 
1172 & 0.0793 & 23 & 670$\pm$130 & 1.60 & 0.497 & 0.41$\pm$0.14 & 0.33$\pm$0.14 & 0.32$\pm$0.13 & 0.29$\pm$0.14 \\ 
3531 & 0.0635 & 17 & 660$\pm$115 & 1.58 & 0.479 & 0.69$\pm$0.19 & 0.80$\pm$0.19 & 0.13$\pm$0.14 & 0.00$\pm$0.14 \\ 
3140 & 0.0449 & 42 & 637$\pm$84 & 1.54 & 0.435 & 0.44$\pm$0.16 & 0.41$\pm$0.17 & 0.56$\pm$0.16 & 0.55$\pm$0.17 \\ 
3050 & 0.0734 & 46 & 635$\pm$67 & 1.52 & 0.424 & 0.71$\pm$0.10 & 0.72$\pm$0.10 & 0.34$\pm$0.11 & 0.32$\pm$0.11 \\ 
1170 & 0.0846 & 23 & 633$\pm$83 & 1.50 & 0.418 & 0.81$\pm$0.09 & 0.85$\pm$0.09 & 0.24$\pm$0.10 & 0.19$\pm$0.09 \\ 
1011 & 0.0847 & 36 & 631$\pm$73 & 1.50 & 0.415 & 0.56$\pm$0.12 & 0.58$\pm$0.12 & 0.33$\pm$0.11 & 0.29$\pm$0.11 \\ 
3123 & 0.0583 & 12 & 629$\pm$146 & 1.51 & 0.416 & 0.36$\pm$0.17 & 0.32$\pm$0.17 & 0.50$\pm$0.18 & 0.54$\pm$0.18 \\ 
2004 & 0.0579 & 93 & 627$\pm$42 & 1.51 & 0.411 & 0.70$\pm$0.08 & 0.71$\pm$0.08 & 0.36$\pm$0.08 & 0.36$\pm$0.08 \\ 
3043 & 0.0698 & 54 & 621$\pm$62 & 1.49 & 0.398 & 0.60$\pm$0.08 & 0.60$\pm$0.08 & 0.29$\pm$0.07 & 0.29$\pm$0.08 \\ 
3478 & 0.0741 & 14 & 617$\pm$75 & 1.47 & 0.388 & 0.45$\pm$0.15 & 0.45$\pm$0.15 & 0.55$\pm$0.15 & 0.56$\pm$0.15 \\ 
1275 & 0.0721 & 18 & 617$\pm$132 & 1.48 & 0.390 & 0.41$\pm$0.14 & 0.37$\pm$0.14 & 0.76$\pm$0.12 & 0.83$\pm$0.12 \\ 
3020 & 0.0629 & 30 & 613$\pm$79 & 1.47 & 0.383 & 0.43$\pm$0.10 & 0.43$\pm$0.10 & 0.39$\pm$0.10 & 0.40$\pm$0.10 \\ 
3319 & 0.0639 & 21 & 601$\pm$40 & 1.44 & 0.361 & 0.35$\pm$0.13 & 0.32$\pm$0.13 & 0.57$\pm$0.13 & 0.59$\pm$0.13 \\ 
3529 & 0.0410 & 19 & 596$\pm$141 & 1.45 & 0.357 & 0.58$\pm$0.19 & 0.55$\pm$0.19 & 0.42$\pm$0.19 & 0.45$\pm$0.19 \\ 
2018 & 0.0550 & 65 & 596$\pm$54 & 1.44 & 0.353 & 0.69$\pm$0.10 & 0.68$\pm$0.10 & 0.13$\pm$0.07 & 0.12$\pm$0.07 \\ 
1017 & 0.0769 & 41 & 596$\pm$56 & 1.42 & 0.350 & 0.83$\pm$0.07 & 0.85$\pm$0.07 & 0.06$\pm$0.05 & 0.04$\pm$0.05 \\ 
1088 & 0.0735 & 32 & 591$\pm$74 & 1.41 & 0.343 & 0.65$\pm$0.11 & 0.71$\pm$0.11 & 0.16$\pm$0.09 & 0.13$\pm$0.09 \\ 
1026 & 0.0720 & 26 & 580$\pm$74 & 1.39 & 0.323 & 0.41$\pm$0.14 & 0.42$\pm$0.14 & 0.59$\pm$0.14 & 0.61$\pm$0.14 \\ 
1024 & 0.0826 & 37 & 579$\pm$89 & 1.38 & 0.320 & 0.50$\pm$0.34 & 0.48$\pm$0.34 & 0.50$\pm$0.34 & 0.45$\pm$0.34 \\ 
3064 & 0.0610 & 42 & 577$\pm$68 & 1.39 & 0.320 & 0.34$\pm$0.10 & 0.33$\pm$0.10 & 0.70$\pm$0.09 & 0.72$\pm$0.10 \\ 
2009 & 0.0530 & 42 & 573$\pm$54 & 1.38 & 0.315 & 0.59$\pm$0.10 & 0.61$\pm$0.11 & 0.36$\pm$0.10 & 0.32$\pm$0.10 \\ 
3275 & 0.0699 & 12 & 570$\pm$133 & 1.36 & 0.307 & 0.32$\pm$0.15 & 0.24$\pm$0.16 & 0.80$\pm$0.13 & 0.90$\pm$0.13 \\ 
3088 & 0.0463 & 61 & 556$\pm$61 & 1.34 & 0.288 & 0.63$\pm$0.12 & 0.63$\pm$0.12 & 0.17$\pm$0.09 & 0.14$\pm$0.09 \\ 
3023 & 0.0719 & 47 & 556$\pm$51 & 1.33 & 0.285 & 0.62$\pm$0.09 & 0.62$\pm$0.10 & 0.18$\pm$0.08 & 0.20$\pm$0.08 \\ 
3041 & 0.0776 & 15 & 555$\pm$72 & 1.32 & 0.283 & 0.45$\pm$0.15 & 0.46$\pm$0.15 & 0.45$\pm$0.15 & 0.45$\pm$0.15 \\ 
2016 & 0.0449 & 135 & 552$\pm$34 & 1.34 & 0.283 & 0.42$\pm$0.09 & 0.41$\pm$0.09 & 0.33$\pm$0.08 & 0.31$\pm$0.08 \\ 
3474 & 0.0519 & 19 & 547$\pm$93 & 1.32 & 0.275 & 0.32$\pm$0.15 & 0.27$\pm$0.16 & 0.44$\pm$0.16 & 0.48$\pm$0.17 \\ 
3505 & 0.0711 & 18 & 543$\pm$96 & 1.30 & 0.266 & 0.55$\pm$0.15 & 0.55$\pm$0.15 & 0.45$\pm$0.15 & 0.50$\pm$0.15 \\ 
3237 & 0.0456 & 30 & 542$\pm$66 & 1.31 & 0.267 & 0.85$\pm$0.10 & 0.90$\pm$0.10 & 0.32$\pm$0.13 & 0.30$\pm$0.14 \\ 
3249 & 0.0740 & 19 & 541$\pm$98 & 1.29 & 0.263 & 0.60$\pm$0.13 & 0.63$\pm$0.13 & 0.47$\pm$0.13 & 0.45$\pm$0.13 \\ 
3117 & 0.0623 & 49 & 536$\pm$52 & 1.29 & 0.257 & 0.72$\pm$0.10 & 0.75$\pm$0.10 & 0.13$\pm$0.07 & 0.08$\pm$0.07 \\ 
3003 & 0.0597 & 59 & 535$\pm$56 & 1.29 & 0.255 & 0.61$\pm$0.09 & 0.62$\pm$0.09 & 0.35$\pm$0.09 & 0.35$\pm$0.09 \\ 
3121 & 0.0627 & 29 & 529$\pm$71 & 1.27 & 0.247 & 0.63$\pm$0.12 & 0.62$\pm$0.12 & 0.30$\pm$0.11 & 0.31$\pm$0.12 \\ 
3167 & 0.0732 & 19 & 526$\pm$82 & 1.26 & 0.240 & 0.70$\pm$0.13 & 0.71$\pm$0.13 & 0.38$\pm$0.13 & 0.37$\pm$0.13 \\ 
2058 & 0.0734 & 21 & 526$\pm$99 & 1.26 & 0.241 & 0.54$\pm$0.14 & 0.52$\pm$0.14 & 0.38$\pm$0.13 & 0.39$\pm$0.14 \\ 
3011 & 0.0820 & 26 & 524$\pm$81 & 1.25 & 0.237 & 0.66$\pm$0.11 & 0.66$\pm$0.11 & 0.24$\pm$0.10 & 0.23$\pm$0.10 \\ 
3015 & 0.0689 & 34 & 523$\pm$67 & 1.25 & 0.238 & 0.59$\pm$0.10 & 0.59$\pm$0.10 & 0.36$\pm$0.10 & 0.35$\pm$0.10 \\ 
2178 & 0.0526 & 14 & 522$\pm$84 & 1.26 & 0.238 & 0.26$\pm$0.17 & 0.21$\pm$0.17 & 0.58$\pm$0.19 & 0.60$\pm$0.19 \\ 
1047 & 0.0829 & 22 & 521$\pm$87 & 1.24 & 0.233 & 0.92$\pm$0.09 & 1.00$\pm$0.09 & 0.08$\pm$0.09 & 0.00$\pm$0.09 \\ 
1120 & 0.0756 & 13 & 516$\pm$109 & 1.23 & 0.227 & 0.44$\pm$0.16 & 0.41$\pm$0.17 & 0.44$\pm$0.16 & 0.44$\pm$0.17 \\ 
1050 & 0.0717 & 14 & 514$\pm$88 & 1.23 & 0.225 & 0.68$\pm$0.13 & 0.65$\pm$0.14 & 0.41$\pm$0.14 & 0.44$\pm$0.14 \\ 
3583 & 0.0673 & 29 & 512$\pm$80 & 1.23 & 0.223 & 0.64$\pm$0.10 & 0.63$\pm$0.10 & 0.31$\pm$0.10 & 0.31$\pm$0.10 \\ 
3196 & 0.0603 & 17 & 512$\pm$92 & 1.23 & 0.224 & 0.29$\pm$0.14 & 0.26$\pm$0.14 & 0.50$\pm$0.16 & 0.51$\pm$0.16 \\ 
2005 & 0.0433 & 48 & 506$\pm$47 & 1.23 & 0.218 & 0.84$\pm$0.11 & 0.91$\pm$0.11 & 0.16$\pm$0.11 & 0.11$\pm$0.11 \\ 
1223 & 0.0600 & 18 & 506$\pm$50 & 1.22 & 0.216 & 0.50$\pm$0.34 & 0.74$\pm$0.34 & 0.50$\pm$0.34 & 0.41$\pm$0.34 \\ 
2020 & 0.0740 & 32 & 505$\pm$83 & 1.21 & 0.213 & 0.68$\pm$0.15 & 0.71$\pm$0.15 & 0.08$\pm$0.09 & 0.00$\pm$0.09 \\ 
1179 & 0.0775 & 16 & 505$\pm$94 & 1.20 & 0.212 & 0.50$\pm$0.14 & 0.45$\pm$0.14 & 0.50$\pm$0.14 & 0.56$\pm$0.14 \\ 
3539 & 0.0542 & 12 & 502$\pm$122 & 1.21 & 0.212 & 0.21$\pm$0.20 & 0.00$\pm$0.20 & 0.50$\pm$0.25 & 0.56$\pm$0.25 \\ 
3080 & 0.0616 & 17 & 501$\pm$110 & 1.20 & 0.209 & 0.68$\pm$0.15 & 0.74$\pm$0.16 & 0.32$\pm$0.15 & 0.25$\pm$0.16 \\ 
2111 & 0.0778 & 8 & 500$\pm$172 & 1.19 & 0.207 & 0.50$\pm$0.34 & 0.77$\pm$0.34 & 0.50$\pm$0.34 & 0.00$\pm$0.34 \\ 
1149 & 0.0523 & 33 & 500$\pm$62 & 1.21 & 0.209 & 0.45$\pm$0.15 & 0.46$\pm$0.15 & 0.55$\pm$0.15 & 0.53$\pm$0.15 \\ 
1404 & 0.0789 & 14 & 498$\pm$155 & 1.19 & 0.204 & 0.44$\pm$0.16 & 0.35$\pm$0.16 & 0.56$\pm$0.16 & 0.66$\pm$0.16 \\ 
1144 & 0.0751 & 20 & 497$\pm$71 & 1.19 & 0.203 & 0.71$\pm$0.14 & 0.71$\pm$0.14 & 0.29$\pm$0.14 & 0.29$\pm$0.14 \\ 
1147 & 0.0809 & 17 & 493$\pm$95 & 1.17 & 0.197 & 0.68$\pm$0.13 & 0.69$\pm$0.13 & 0.15$\pm$0.10 & 0.14$\pm$0.10 \\ 
3019 & 0.0682 & 32 & 492$\pm$67 & 1.18 & 0.198 & 0.74$\pm$0.11 & 0.78$\pm$0.11 & 0.53$\pm$0.13 & 0.54$\pm$0.13 \\ 
1178 & 0.0543 & 32 & 490$\pm$41 & 1.18 & 0.197 & 0.54$\pm$0.14 & 0.51$\pm$0.14 & 0.54$\pm$0.14 & 0.60$\pm$0.13 \\ 
3066 & 0.0767 & 30 & 482$\pm$49 & 1.15 & 0.185 & 0.61$\pm$0.10 & 0.62$\pm$0.10 & 0.30$\pm$0.09 & 0.24$\pm$0.09 \\ 
2027 & 0.0790 & 34 & 481$\pm$62 & 1.14 & 0.184 & 0.50$\pm$0.34 & 0.54$\pm$0.34 & 0.50$\pm$0.34 & 0.18$\pm$0.34 \\ 
3344 & 0.0738 & 8 & 479$\pm$63 & 1.14 & 0.182 & 0.80$\pm$0.13 & 0.78$\pm$0.16 & 0.32$\pm$0.15 & 0.35$\pm$0.16 \\ 
3083 & 0.0760 & 19 & 470$\pm$64 & 1.12 & 0.172 & 0.50$\pm$0.13 & 0.53$\pm$0.13 & 0.50$\pm$0.13 & 0.48$\pm$0.13 \\ 
1087 & 0.0782 & 15 & 465$\pm$180 & 1.11 & 0.166 & 0.37$\pm$0.12 & 0.37$\pm$0.12 & 0.63$\pm$0.12 & 0.60$\pm$0.12 \\ 
1005 & 0.0809 & 21 & 458$\pm$50 & 1.09 & 0.159 & 0.50$\pm$0.20 & 0.50$\pm$0.20 & 0.31$\pm$0.19 & 0.25$\pm$0.19 \\ 
1360 & 0.0746 & 8 & 448$\pm$103 & 1.07 & 0.149 & 0.23$\pm$0.15 & 0.18$\pm$0.15 & 0.77$\pm$0.15 & 0.84$\pm$0.15 \\ 
3120 & 0.0731 & 19 & 445$\pm$80 & 1.06 & 0.146 & 0.70$\pm$0.13 & 0.70$\pm$0.13 & 0.38$\pm$0.13 & 0.38$\pm$0.13 \\ 
3248 & 0.0446 & 20 & 442$\pm$90 & 1.07 & 0.145 & 0.26$\pm$0.17 & 0.27$\pm$0.17 & 0.58$\pm$0.19 & 0.56$\pm$0.19 \\ 
3069 & 0.0555 & 20 & 441$\pm$47 & 1.06 & 0.144 & 0.76$\pm$0.12 & 0.82$\pm$0.12 & 0.41$\pm$0.14 & 0.39$\pm$0.14 \\ 
2121 & 0.0817 & 13 & 441$\pm$86 & 1.05 & 0.141 & 0.42$\pm$0.19 & 0.37$\pm$0.19 & 0.58$\pm$0.19 & 0.63$\pm$0.19 \\ 
2141 & 0.0779 & 16 & 422$\pm$89 & 1.00 & 0.124 & 0.80$\pm$0.13 & 0.87$\pm$0.13 & 0.20$\pm$0.13 & 0.11$\pm$0.13 \\ 
3506 & 0.0635 & 11 & 421$\pm$87 & 1.01 & 0.124 & 0.26$\pm$0.17 & 0.22$\pm$0.17 & 0.74$\pm$0.17 & 0.79$\pm$0.17 \\ 
1368 & 0.0520 & 10 & 417$\pm$91 & 1.01 & 0.121 & 0.61$\pm$0.22 & 0.72$\pm$0.22 & 0.39$\pm$0.22 & 0.28$\pm$0.22 \\ 
2052 & 0.0772 & 12 & 415$\pm$72 & 0.99 & 0.118 & 0.50$\pm$0.14 & 0.49$\pm$0.14 & 0.15$\pm$0.10 & 0.10$\pm$0.10 \\ 
2000 & 0.0699 & 29 & 411$\pm$53 & 0.98 & 0.115 & 0.48$\pm$0.10 & 0.48$\pm$0.10 & 0.43$\pm$0.10 & 0.43$\pm$0.10 \\ 
1009 & 0.0746 & 27 & 404$\pm$41 & 0.97 & 0.110 & 0.53$\pm$0.11 & 0.52$\pm$0.11 & 0.31$\pm$0.11 & 0.33$\pm$0.11 \\ 
3146 & 0.0459 & 24 & 403$\pm$45 & 0.97 & 0.110 & 0.55$\pm$0.15 & 0.53$\pm$0.15 & 0.07$\pm$0.08 & 0.00$\pm$0.08 \\ 
1043 & 0.0743 & 30 & 403$\pm$60 & 0.96 & 0.109 & 0.74$\pm$0.11 & 0.79$\pm$0.11 & 0.26$\pm$0.11 & 0.22$\pm$0.11 \\ 
1106 & 0.0402 & 18 & 402$\pm$112 & 0.98 & 0.109 & 0.39$\pm$0.15 & 0.34$\pm$0.15 & 0.50$\pm$0.16 & 0.44$\pm$0.16 \\ 
3572 & 0.0587 & 11 & 393$\pm$183 & 0.95 & 0.102 & 0.79$\pm$0.20 & 1.00$\pm$0.20 & 0.21$\pm$0.20 & 0.00$\pm$0.20 \\ 
3144 & 0.0805 & 16 & 386$\pm$57 & 0.92 & 0.095 & 0.94$\pm$0.06 & 1.00$\pm$0.06 & 0.14$\pm$0.09 & 0.08$\pm$0.10 \\ 
2010 & 0.0645 & 18 & 384$\pm$55 & 0.92 & 0.095 & 0.50$\pm$0.18 & 0.46$\pm$0.18 & 0.36$\pm$0.17 & 0.36$\pm$0.17 \\ 
2211 & 0.0422 & 22 & 372$\pm$38 & 0.90 & 0.087 & 0.68$\pm$0.15 & 0.74$\pm$0.16 & 0.32$\pm$0.15 & 0.31$\pm$0.16 \\ 
1341 & 0.0700 & 10 & 370$\pm$59 & 0.88 & 0.084 & 0.58$\pm$0.19 & 0.56$\pm$0.19 & 0.26$\pm$0.17 & 0.21$\pm$0.17 \\ 
3388 & 0.0722 & 12 & 357$\pm$109 & 0.85 & 0.075 & 0.61$\pm$0.22 & 0.66$\pm$0.22 & 0.61$\pm$0.22 & 0.63$\pm$0.22 \\ 
3007 & 0.0685 & 32 & 356$\pm$50 & 0.85 & 0.075 & 0.73$\pm$0.09 & 0.74$\pm$0.09 & 0.17$\pm$0.08 & 0.13$\pm$0.08 \\ 
2140 & 0.0810 & 5 & 355$\pm$216 & 0.84 & 0.074 & 0.13$\pm$0.14 & 0.00$\pm$0.14 & 0.87$\pm$0.14 & 1.00$\pm$0.14 \\ 
3182 & 0.0614 & 18 & 352$\pm$84 & 0.85 & 0.073 & 0.44$\pm$0.16 & 0.42$\pm$0.17 & 0.20$\pm$0.13 & 0.16$\pm$0.13 \\ 
1006 & 0.0477 & 24 & 340$\pm$54 & 0.82 & 0.066 & 0.77$\pm$0.15 & 0.85$\pm$0.15 & 0.23$\pm$0.15 & 0.15$\pm$0.15 \\ 
2120 & 0.0518 & 12 & 338$\pm$101 & 0.82 & 0.065 & 0.58$\pm$0.19 & 0.58$\pm$0.19 & 0.42$\pm$0.19 & 0.41$\pm$0.19 \\ 
3341 & 0.0723 & 14 & 331$\pm$97 & 0.79 & 0.060 & 0.56$\pm$0.16 & 0.55$\pm$0.17 & 0.32$\pm$0.15 & 0.30$\pm$0.16 \\ 
3272 & 0.0692 & 9 & 329$\pm$54 & 0.79 & 0.059 & 0.69$\pm$0.19 & 0.72$\pm$0.19 & 0.50$\pm$0.20 & 0.55$\pm$0.20 \\ 
2156 & 0.0668 & 17 & 328$\pm$50 & 0.79 & 0.059 & 0.82$\pm$0.12 & 0.89$\pm$0.12 & 0.18$\pm$0.12 & 0.11$\pm$0.12 \\ 
1015 & 0.0792 & 10 & 327$\pm$66 & 0.78 & 0.058 & 0.79$\pm$0.20 & 1.00$\pm$0.20 & 0.50$\pm$0.25 & 0.57$\pm$0.25 \\ 
2122 & 0.0826 & 7 & 315$\pm$61 & 0.75 & 0.052 & 0.42$\pm$0.19 & 0.31$\pm$0.19 & 0.42$\pm$0.19 & 0.31$\pm$0.19 \\ 
3340 & 0.0611 & 8 & 314$\pm$60 & 0.75 & 0.051 & 0.79$\pm$0.20 & 1.00$\pm$0.20 & 0.21$\pm$0.20 & 0.00$\pm$0.20 \\ 
3247 & 0.0571 & 14 & 307$\pm$97 & 0.74 & 0.048 & 0.29$\pm$0.26 & 0.00$\pm$0.26 & 0.71$\pm$0.26 & 1.00$\pm$0.26 \\ 
3061 & 0.0480 & 21 & 300$\pm$37 & 0.72 & 0.045 & 0.58$\pm$0.19 & 0.58$\pm$0.19 & 0.42$\pm$0.19 & 0.37$\pm$0.19 \\ 
1040 & 0.0539 & 16 & 299$\pm$38 & 0.72 & 0.045 & 0.39$\pm$0.22 & 0.40$\pm$0.22 & 0.39$\pm$0.22 & 0.36$\pm$0.22 \\ 
3267 & 0.0710 & 5 & 298$\pm$147 & 0.71 & 0.044 & 0.31$\pm$0.19 & 0.20$\pm$0.19 & 0.50$\pm$0.20 & 0.50$\pm$0.20 \\ 
3328 & 0.0412 & 12 & 297$\pm$40 & 0.72 & 0.044 & 0.50$\pm$0.25 & 0.54$\pm$0.25 & 0.50$\pm$0.25 & 0.43$\pm$0.25 \\ 
2070 & 0.0775 & 12 & 295$\pm$52 & 0.70 & 0.043 & 0.56$\pm$0.16 & 0.57$\pm$0.17 & 0.44$\pm$0.16 & 0.42$\pm$0.17 \\ 
3169 & 0.0756 & 7 & 291$\pm$146 & 0.70 & 0.041 & 0.32$\pm$0.15 & 0.30$\pm$0.16 & 0.80$\pm$0.13 & 0.91$\pm$0.13 \\ 
3262 & 0.0452 & 32 & 282$\pm$48 & 0.68 & 0.038 & 0.61$\pm$0.22 & 0.67$\pm$0.22 & 0.61$\pm$0.22 & 0.74$\pm$0.22 \\ 
1297 & 0.0451 & 10 & 271$\pm$85 & 0.66 & 0.033 & 0.39$\pm$0.22 & 0.27$\pm$0.22 & 0.39$\pm$0.22 & 0.33$\pm$0.22 \\ 
3147 & 0.0534 & 11 & 265$\pm$80 & 0.64 & 0.031 & 0.69$\pm$0.19 & 0.82$\pm$0.19 & 0.13$\pm$0.14 & 0.00$\pm$0.14 \\ 
1128 & 0.0619 & 12 & 257$\pm$64 & 0.62 & 0.028 & 0.64$\pm$0.17 & 0.66$\pm$0.17 & 0.36$\pm$0.17 & 0.35$\pm$0.17 \\ 
3062 & 0.0615 & 12 & 256$\pm$44 & 0.62 & 0.028 & 0.56$\pm$0.16 & 0.55$\pm$0.17 & 0.44$\pm$0.16 & 0.45$\pm$0.17 \\ 
3318 & 0.0419 & 7 & 252$\pm$56 & 0.61 & 0.027 & 0.50$\pm$0.20 & 0.48$\pm$0.21 & 0.50$\pm$0.20 & 0.51$\pm$0.20 \\ 
1171 & 0.0556 & 15 & 242$\pm$48 & 0.58 & 0.024 & 0.61$\pm$0.22 & 0.67$\pm$0.22 & 0.39$\pm$0.22 & 0.33$\pm$0.22 \\ 
1384 & 0.0541 & 7 & 240$\pm$55 & 0.58 & 0.023 & 0.71$\pm$0.26 & 1.00$\pm$0.26 & 0.29$\pm$0.26 & 0.00$\pm$0.26 \\ 
3581 & 0.0591 & 5 & 236$\pm$57 & 0.57 & 0.022 & 0.50$\pm$0.25 & 0.38$\pm$0.25 & 0.79$\pm$0.20 & 1.00$\pm$0.20 \\ 
3407 & 0.0652 & 4 & 233$\pm$107 & 0.56 & 0.021 & 0.29$\pm$0.26 & 0.00$\pm$0.26 & 0.29$\pm$0.26 & 0.00$\pm$0.26 \\ 
2150 & 0.0415 & 9 & 221$\pm$66 & 0.53 & 0.018 & 0.50$\pm$0.34 & 0.13$\pm$0.34 & 0.50$\pm$0.34 & 0.71$\pm$0.34 \\ 
2112 & 0.0510 & 12 & 212$\pm$59 & 0.51 & 0.016 & 0.71$\pm$0.26 & 1.00$\pm$0.26 & 0.29$\pm$0.26 & 0.00$\pm$0.26 \\ 
3553 & 0.0434 & 10 & 208$\pm$53 & 0.50 & 0.015 & 0.61$\pm$0.22 & 0.67$\pm$0.22 & 0.61$\pm$0.22 & 0.63$\pm$0.22 \\ 
3195 & 0.0599 & 11 & 206$\pm$31 & 0.50 & 0.015 & 0.69$\pm$0.19 & 0.72$\pm$0.19 & 0.50$\pm$0.20 & 0.60$\pm$0.21 \\ 
3052 & 0.0746 & 7 & 202$\pm$77 & 0.48 & 0.014 & 0.50$\pm$0.20 & 0.47$\pm$0.20 & 0.31$\pm$0.19 & 0.27$\pm$0.19 \\ 
1344 & 0.0747 & 7 & 202$\pm$47 & 0.48 & 0.014 & 0.77$\pm$0.15 & 0.84$\pm$0.15 & 0.50$\pm$0.18 & 0.56$\pm$0.18 \\ 
3428 & 0.0730 & 5 & 193$\pm$41 & 0.46 & 0.012 & 0.39$\pm$0.22 & 0.34$\pm$0.22 & 0.61$\pm$0.22 & 0.65$\pm$0.22 \\ 
3345 & 0.0516 & 4 & 142$\pm$57 & 0.34 & 0.005 & 0.71$\pm$0.26 & 1.00$\pm$0.26 & 0.29$\pm$0.26 & 0.00$\pm$0.26 \\ 
3265 & 0.0597 & 4 & 142$\pm$11 & 0.34 & 0.005 & 0.50$\pm$0.25 & 0.43$\pm$0.25 & 0.50$\pm$0.25 & 0.43$\pm$0.25 \\ 
3355 & 0.0677 & 4 & 125$\pm$61 & 0.30 & 0.003 & 0.84$\pm$0.16 & 1.00$\pm$0.16 & 0.16$\pm$0.16 & 0.00$\pm$0.16 \\ 
3244 & 0.0600 & 4 & 120$\pm$20 & 0.29 & 0.003 & 0.71$\pm$0.26 & 1.00$\pm$0.26 & 0.71$\pm$0.26 & 1.00$\pm$0.26
\footnotetext[3]{From \cite{linden07}}
\footnotetext[4]{Number of spectroscopically-confirmed cluster members from \citet{linden07}}
\end{longtable}
}

\end{document}